\documentstyle[12pt]{article}

\begin{document}
\renewcommand{\thefootnote}{\fnsymbol{footnote}}
\begin{titlepage}
\begin{flushright}
hep-th/0612284
\end{flushright}

\vspace{10mm}
\begin{center}
{\Large\bf New Born-Infeld and $Dp$-Brane Actions under 2-Metric and 3-Metric Prescriptions}
\vspace{25mm}

{\large
Yan-Gang Miao\footnote{Email addresses:
 miaoyg@nankai.edu.cn; ymiao@icpt.it}}\\
\vspace{10mm} $^{1)}${\em Department of Physics, Nankai University, Tianjin 300071, \\
People's Republic of China\footnote{Permanent address.}}

$^{2)}${\em The Abdus Salam International Centre for Theoretical Physics,\\
Strada Costiera 11, 34014 Trieste, Italy}

\end{center}
\vspace{15mm}
\centerline{{\bf{Abstract}}}
\vspace{5mm}

The parent action method is utilized to the Born-Infeld and $Dp$-brane theories.
Various new forms of Born-Infeld and $Dp$-brane actions are derived by using this systematic approach,
in which both the already known 2-metric and newly proposed 3-metric prescriptions are considered.
An auxiliary worldvolume tensor field, denoted by ${\omega}_{{\mu}{\nu}}$, is introduced and treated probably as an additional worldvolume metric
because it plays a similar role to that of the auxiliary worldvolume
(also called {\em intrinsic}) metric ${\gamma}_{{\mu}{\nu}}$. Some properties, such as duality, permutation and Weyl invariance as a local
worldvolume symmetry of the new forms
are analyzed. In particular, a new symmetry, i.e. the double Weyl invariance is discovered in 3-metric forms.

\end{titlepage}
\newpage
\renewcommand{\thefootnote}{\arabic{footnote}}
\setcounter{footnote}{0}
\setcounter{page}{2}

\section{Introduction}

The remarkable progress of string theory~\cite{s1} is the discovery of Dirichlet $p$-branes~\cite{s2},
i.e. $Dp$-branes. Geometrically,
the $Dp$-branes are $(p+1)$-dimensional hypersurfaces that are embedded in
a higher dimensional spacetime. Dynamically, they are solitonic solutions to string equations that are ``branes'' on which open strings attach
with Dirichlet boundary conditions. The dynamics of $Dp$-branes is induced by the open strings and governed in general by
the action of Born-Infeld type~\cite{s3}. Recently, many different actions for generalizations of the Born-Infeld have been
proposed to describe the effective worldvolume theories of $Dp$-branes. For instance, see refs.~\cite{s4,s5,s6,s7,s8,s9,s10,s11,s12,s13,s14,s15,s16,s17}
where quite interesting are the action form~\cite{s14} that is quadratic in abelian field strengths\footnote{Different quadratic formulations
with non-symmetric auxiliary worldvolume metrics are also given by refs.~\cite{s13,s16}. However, they do not
have such a permutation symmetry between $g_{{\mu}{\nu}}$ and ${\cal G}_{{\mu}{\nu}}$ (cf. eq.~(3)) that is dealt with as our starting point and
are therefore not discussed in this paper.}
and its conformal invariant development~\cite{s15}. The merit of this formalism, as promised and expected by refs.~\cite{s14,s15}, is on simplifying
quantization and
dualization of the gauge fields, which originates from the fact that the string action~\cite{s18} with an auxiliary worldsheet metric and conformal
invariance greatly simplifies the analysis of string theory and allows a covariant quantization~\cite{s19}.

The idea of the parent action method was introduced~\cite{s20} in order to establish, at the level of
lagrangians instead of equations of motion, the equivalence or so-called
duality between the abelian self-dual and Maxwell-Chern-Simons models in
(2+1)-dimensional spacetime. Recently the method has been developed and applied to a quite wide region.
For instance, one direct development~\cite{s21} is the building of the duality between the non-abelian
self-dual and Yang-Mills-Chern-Simons models.
One interesting application related closely to the present paper focuses on chiral bosons~\cite{s22} and bosonic $p$-branes~\cite{s23}.
For chiral bosons and their extensions to chiral $p$-forms, the self-duality of various chiral boson actions has been built~\cite{s24}
with the modification~\cite{s25} of
the method with one more auxiliary field to preserve the manifest Lorentz invariance of the actions. For bosonic $p$-branes,
some new actions of bosonic $p$-branes have been worked out and their classification to dual sets fixed~\cite{s26} as well, and furthermore the canonical
Hamiltonian analyses of the new actions have therefore been performed~\cite{s27}. In the present work, the term duality is used to refer to different
actions for the same $Dp$-brane, rather than the duality of the $Dp$-branes themselves.

The main idea of the parent action method~\cite{s28} originates from the
Legendre transformation and contains the meaning of the two aspects:
(a) to introduce auxiliary fields and then construct a parent or master
action by adding a Lagrange multiplier term to a known action, and (b) to make variation of the parent
action with respect to each auxiliary field, solve one auxiliary field
in terms of other fields and then substitute the solution into the parent
action. Through making variations with respect to different auxiliary fields,
we can obtain different forms of the actions. The actions are, of
course, equivalent classically, and the relation between them is usually
referred to as duality. If the resulting actions are the same,
the relation is called self-duality.

The content and arrangement of this paper are as follows. In the next section, we begin to construct 2-metric forms
in terms of the parent action
method, and then build, as a by-product of the method, duality structures of the action forms.
In the application of the method, we consider two proposals of writing parent actions one of which,
introduced firstly by the present author~\cite{s26}, has given rise to interesting
bosonic $p$-brane actions with highly nonlinear terms of induced metrics~\cite{s27}.
Because of some symmetric formalism of the 1-metric source action
(see section 2 for details) we start with, we obtain a variety of new actions of $Dp$-branes (especially for 3-metric forms,
see section 3 for details). Although some forms
do not provide useful formulations for simplifying the analysis of the $Dp$-brane theory, such as ${S^{\prime}}^{{\rm II}}$ (cf. eqs.~(18)--(23))
that are not quadratic in field strengths,
they supply the base for us to
go further beyond the actions with 2-metrics.
This is just the task of the following section. In section 3, we propose the 3-metric prescription whose key is the introduction
of an additional worldvolume metric. Simply speaking, the motivation\footnote{This idea is necessary for the investigation of $Dp$-branes in general
though the typical 2-metric action~\cite{s14}, when it describes the low-energy effective worldvolume theory for an open type I string, can be reduced to
its static gauge form which is quadratic in the derivatives of both gauge fields and spacetime coordinates.}
is to look for such a $Dp$-brane action that is, besides quadratic in field strengths, a rational functional of the induced
metric. It is obvious that a rational formulation is more convenient to be analyzed than a rooted one.
Using the parent action method two successive times to the 1-metric source form, together with the consideration of the two proposals
just mentioned, we get a large amount of new $Dp$-brane actions with 3-metrics and also fix their duality structures. Quite surprising is the richness
of the 3-metric forms in permutation and conformal symmetry. We therefore devote section 4 to analyze the properties in detail.
We give an interesting one-to-one correspondence of the 3-metric actions under permutation transformations of the two auxiliary worldvolume metrics,
and, in particular, discover
a new symmetry, i.e. the double conformal invariance in some 3-metric action forms.
Finally, a conclusion is made in section 5.

In this paper we name various forms of actions at first by using series number that is just the number of metrics involved in, and then classify them
within series {\rm II} and series {\rm III} (series {\rm I} includes only one form) in terms of some important properties they have, such as duality,
permutation and Weyl symmetries.

The notation we use throughout this paper is as follows.
Some Greek lowercase letters, for example, ${\mu},{\nu},{\lambda},{\sigma}$, running over $0,1,\cdots,p$, are used as indices in the worldvolume
that is spanned by $p+1$ arbitrary parameters ${\xi}^{\mu}$. Incidentally, spacetime indices are suppressed because our discussions only involve in
the worldvolume.
The $Dp$-brane kinetic term takes the form~\cite{s2}
\begin{equation}
S=-T_{p}\int d^{p+1}{\xi}e^{-\phi}\sqrt{-{\rm det}(g_{{\mu}{\nu}}+{\cal F}_{{\mu}{\nu}})},
\end{equation}
where
\begin{equation}
{\cal F}_{{\mu}{\nu}}\equiv F_{{\mu}{\nu}}-B_{{\mu}{\nu}},
\end{equation}
$\phi$, $g_{{\mu}{\nu}}$ and $B_{{\mu}{\nu}}$ are pullbacks to the worldvolume of the background dilaton, metric and NS antisymmetric two-form fields,
and $F_{{\mu}{\nu}}={\partial}_{\mu}A_{\nu}-{\partial}_{\nu}A_{\mu}$, with $A_{\mu}({\xi})$ the $U(1)$ worldvolume gauge field.
$T_{p}$ is the $Dp$-brane tension. Eq.~(1) can be rewritten as~\cite{s14}
\begin{equation}
S^{{\rm I}}=-T_{p}\int d^{p+1}{\xi}e^{-\phi}(-g)^{1/4}(-{\cal G})^{1/4},
\end{equation}
where
\begin{equation}
{\cal G}_{{\mu}{\nu}}=g_{{\mu}{\nu}}-g^{{\lambda}{\sigma}}{\cal F}_{{\mu}{\lambda}}{\cal F}_{{\sigma}{\nu}},
\end{equation}
and $g\equiv {\rm det}(g_{{\mu}{\nu}})$, ${\cal G}\equiv {\rm det}({\cal G}_{{\mu}{\nu}})$; $g^{{\mu}{\nu}}$ is the inverse of $g_{{\mu}{\nu}}$. This is
a convenient form for our following discussions, and
treated as the source form of $Dp$-brane actions
we start with in this paper. It is named as 1-metric series with the superscript {\rm I}, i.e. series {\rm I},
which contains solely the induced metric $g_{{\mu}{\nu}}$.

\section{2-Metric Form and Duality}

Starting from the 1-metric form of $Dp$-brane actions, we now derive new forms with 2-metrics, called  2-metric series or series {\rm II},
in terms of the parent action method which has been
shown~\cite{s26}
powerful to discovering new actions and fixing their dualities for bosonic $p$-branes. We point out that this method is also applicable to
Born-Infeld and $Dp$-brane theories and that new forms of actions will be obtained in a systematic way. This means that we obtain not only the forms
proposed already~\cite{s14,s15} but also some new forms unknown before. As the method has been utilized in detail to
bosonic $p$-branes in our previous work~\cite{s26}, here we just follow the main procedure of the method and write down results.

According to the parent action method~\cite{s20,s28}, we introduce
two auxiliary worldvolume second-rank tensor fields ${\Lambda}^{{\mu}{\nu}}$ and ${\gamma}_{{\mu}{\nu}}$, where ${\Lambda}^{{\mu}{\nu}}$ is dealt with as
a Lagrange multiplier,
and write down one parent action of the 1-metric form
\begin{equation}
S^{{\rm I}}_{\rm P1}=-T_{p}\int d^{p+1}{\xi}e^{-\phi}\left[(-g)^{1/4}(-{\gamma})^{1/4}
+{\Lambda}^{{\mu}{\nu}}\left({\gamma}_{{\mu}{\nu}}-{\cal G}_{{\mu}{\nu}}\right)\right],
\end{equation}
where ${\gamma}\equiv {\rm det}({\gamma}_{{\mu}{\nu}})$. Note that the Lagrange multiplier term in the above equation is composed of the contravariant
${\Lambda}^{{\mu}{\nu}}$ multiplied by the covariant ${\gamma}_{{\mu}{\nu}}-{\cal G}_{{\mu}{\nu}}$. This is the first proposal we suggest for
construction of
parent actions while the second, just opposite to the first, will be given later in this section.
We will find that ${\gamma}_{{\mu}{\nu}}$ is
just the auxiliary worldvolume ({\em intrinsic}) metric, but at present it is treated as
an independent auxiliary field.

Now varying eq.~(5) with respect to ${\Lambda}^{{\mu}{\nu}}$ gives the relation
${\gamma}_{{\mu}{\nu}}={\cal G}_{{\mu}{\nu}}$, together with which eq.~(5) turns back to the 1-metric
form eq.~(3). This shows the classical equivalence between the parent and
1-metric actions. However, varying eq.~(5) with respect to ${\gamma}_{{\mu}{\nu}}$
leads to the expression of ${\Lambda}^{{\mu}{\nu}}$ in terms of ${\gamma}_{{\mu}{\nu}}$:
\begin{equation}
{\Lambda}^{{\mu}{\nu}}=-\frac{1}{4}(-g)^{1/4}(-{\gamma})^{1/4}{\gamma}^{{\mu}{\nu}},
\end{equation}
where ${\gamma}^{{\mu}{\nu}}$ is the inverse of ${\gamma}_{{\mu}{\nu}}$. Substituting eq.~(6) into eq.~(5),
we obtain one dual version of the 1-metric action
\begin{equation}
S^{{\rm II}}_{1}=-\frac{T_{p}}{4}\int d^{p+1}{\xi}e^{-\phi}(-g)^{1/4}(-{\gamma})^{1/4}\left[{\gamma}^{{\mu}{\nu}}{\cal G}_{{\mu}{\nu}}-(p-3)
\right].
\end{equation}
This is the 2-metric $Dp$-brane action with the auxiliary field ${\gamma}_{{\mu}{\nu}}$ that now
plays the role of the auxiliary worldvolume metric. It was obtained but its duality to the 1-metric form was not uncovered in ref.~\cite{s14}.
The superscript {\rm II} of the symbol in eq.~(7) means that the action belongs to series
{\rm II}, i.e. the forms with 2-metrics
and the subscript $i$ (here $i=1$) corresponds to the forms that are derived in terms of the first proposal of writing parent actions
mentioned above.

As $S^{{\rm II}}_{1}$ has no Weyl invariance for the general case of $p\neq 3$, we then adopt the approach~\cite{s16,s26}
utilized in bosonic $p$-branes
to derive other 2-metric $Dp$-brane actions that possess such an invariance. To this end, by introducing an auxiliary scalar field ${\Phi}({\xi})$,
and rescaling
the worldvolume metric ${\gamma}_{{\mu}{\nu}}\to{\Phi}{\gamma}_{{\mu}{\nu}}$ in the parent
action of the 1-metric form, i.e. in eq.~(5), we write down the second parent action\footnote{Parent actions are not unique.}
\begin{equation}
S^{{\rm I}}_{\rm P2}=-T_{p}\int d^{p+1}{\xi}e^{-\phi}\left[{\Phi}^{(p+1)/4}(-g)^{1/4}(-{\gamma})^{1/4}
+{\Lambda}^{{\mu}{\nu}}\left({\Phi}{\gamma}_{{\mu}{\nu}}-{\cal G}_{{\mu}{\nu}}\right)\right],
\end{equation}
where ${\Phi}({\xi})$ should be a scalar in both the spacetime and
worldvolume in order to keep eq.~(8) invariant under the Lorentz
transformation and reparametrization.

Varying eq.~(8) with respect to ${\Lambda}^{{\mu}{\nu}}$ brings about
${\gamma}_{{\mu}{\nu}}={\Phi}^{-1}{\cal G}_{{\mu}{\nu}}$, which leads to nothing new but the classical
equivalence between the 1-metric and second parent actions. However,
varying the equation with respect to ${\gamma}_{{\mu}{\nu}}$, we solve ${\Lambda}^{{\mu}{\nu}}$ as follows:
\begin{equation}
{\Lambda}^{{\mu}{\nu}}=-\frac{1}{4}{\Phi}^{(p-3)/4}(-g)^{1/4}(-{\gamma})^{1/4}{\gamma}^{{\mu}{\nu}}.
\end{equation}
Substituting eq.~(9) back to eq.~(8), we derive one more dual action of the 1-metric form
\begin{equation}
S^{{\rm II}}_{2}=-\frac{T_{p}}{4}\int d^{p+1}{\xi}e^{-\phi}(-g)^{1/4}(-{\gamma})^{1/4}\left[{\Phi}^{(p-3)/4}{\gamma}^{{\mu}{\nu}}
{\cal G}_{{\mu}{\nu}}-(p-3){\Phi}^{(p+1)/4}
\right],
\end{equation}
which was obtained too but whose duality to $S^{{\rm I}}$ was not uncovered either in ref.~\cite{s15}.
This action is interesting because it has Weyl invariance that will be analyzed in detail in section 4.

Moreover, we are able to deduce from $S^{{\rm II}}_{2}$ another Weyl invariant form without the auxiliary scalar field ${\Phi}({\xi})$. To this end,
let us vary eq.~(10) with respect to ${\Phi}({\xi})$, which gives rise to the relation for the general case of $p\neq 3$:
\begin{equation}
{\Phi}({\xi})=\frac{1}{p+1}{\gamma}^{{\mu}{\nu}}{\cal G}_{{\mu}{\nu}}.
\end{equation}
Substituting eq.~(11) into eq.~(10), we therefore have the Weyl invariant 2-metric action that does not involve in ${\Phi}({\xi})$
but contains higher
(than two) order terms of field strengths for $Dp$-branes of $p > 3$,
\begin{equation}
S^{{\rm II}}_{3}=-T_{p}\int d^{p+1}{\xi}e^{-\phi}(-g)^{1/4}(-{\gamma})^{1/4}\left(\frac{1}{p+1}{\gamma}^{{\mu}{\nu}}{\cal G}_{{\mu}{\nu}}
\right)^{(p+1)/4}.
\end{equation}
As to the special case of $p=3$, ${\Phi}({\xi})$ disappears automatically from $S^{{\rm II}}_{2}$. Furthermore, we find that
$S^{{\rm II}}_{i}$ ($i=1,2,3$) coincide with each other in this case.  Eq.~(12) is therefore suitable for describing $D3$-branes. Incidentally, a similar
situation also happened to bosonic strings ($1$-branes)~\cite{s23}.

Besides the above two parent actions, we are able to write down according to the first proposal other different forms of parent actions
that associate with the 1-metric source action if we follow the way of dealing with bosonic $p$-branes. As the procedure is quite straightforward to
$Dp$-branes, we here omit it and simply conclude that no more actions are derived
but more dualities among $S^{{\rm I}}$ and $S^{{\rm II}}_{i}$ ($i=1,2,3$) that consist of a closed set of dual actions are exposed. The dualities
are shown in detail by the schematic representation of Figure 1.

At the present stage, although we uncover the dualities that exist in the 1-metric and three 2-metric forms, we merely reproduce
through the parent action method the 2-metric $Dp$-brane actions, $S^{{\rm II}}_{i}$ ($i=1,2,3$),
that have been proposed~\cite{s14,s15} with the consideration different from ours. As we pointed out before that the parent action method is powerful to
finding new action forms and building their dualities, our next goal is to derive other 2-metric forms. To this end, let us introduce
the second proposal for writing parent actions which has been exploited to bosonic $p$-branes~\cite{s26}. Contrary to it in
eq.~(5), now the Lagrange multiplier term  is composed of the covariant
${\Lambda}_{{\mu}{\nu}}$ multiplied by the contravariant ${\gamma}^{{\mu}{\nu}}-{\cal G}^{{\mu}{\nu}}$, where ${\Lambda}_{{\mu}{\nu}}$
and ${\gamma}^{{\mu}{\nu}}$ are two auxiliary worldvolume second-rank tensor fields we introduce, and ${\cal G}^{{\mu}{\nu}}$ is the inverse of
${\cal G}_{{\mu}{\nu}}$. Note that ${\cal G}^{{\mu}{\nu}}$ contain of course highly nonlinear terms of field strengths. As a result, we construct
one parent action that associates with the second proposal:
\begin{equation}
S^{{\rm I}}_{\rm P{\bar 1}}=-T_{p}\int d^{p+1}{\xi}e^{-\phi}\left[(-g)^{1/4}(-{\gamma})^{1/4}
+{\Lambda}_{{\mu}{\nu}}\left({\gamma}^{{\mu}{\nu}}-{\cal G}^{{\mu}{\nu}}\right)\right],
\end{equation}
where subscript ${\bar i}$ (here ${\bar i}={\bar 1}$) corresponds to our second proposal. At first sight $S^{{\rm I}}_{\rm P{\bar 1}}$ looks like
$S^{{\rm I}}_{\rm P1}$ in eq.~(5), just with
the exchange of superscripts and subscripts in the second term. We note that
if ${\cal G}^{{\mu}{\nu}}$ were understood as ${\gamma}^{{\mu}{\lambda}}{\gamma}^{{\nu}{\sigma}}{\cal G}_{{\lambda}{\sigma}}$,
$S^{{\rm I}}_{\rm P{\bar 1}}$ was
exactly the same as $S^{{\rm I}}_{\rm P1}$ and nothing new could be deduced from
eq.~(13) but the $Dp$-brane action eq.~(7). Actually ${\cal G}^{{\mu}{\nu}}$ in
eq.~(13) is defined as the inverse of ${\cal G}_{{\mu}{\nu}}$ and is thus independent of
${\gamma}_{{\mu}{\nu}}$. With this in mind, we follow the usual procedure described
above and derive a new 2-metric $Dp$-brane action that is dual and equivalent
to $S^{{\rm I}}$:
\begin{equation}
S^{{\rm II}}_{\bar{1}}=-\frac{T_{p}}{4}\int d^{p+1}{\xi}e^{-\phi}(-g)^{1/4}(-{\gamma})^{1/4}\left[{-\gamma}_{{\mu}{\nu}}{\cal G}^{{\mu}{\nu}}+(p+5)
\right].
\end{equation}
The new $Dp$-brane action is Lorentz invariant and reparametrization
invariant as well. Under the reparametrization, the factor
$d^{p+1}{\xi}e^{-\phi}(-g)^{1/4}(-{\gamma})^{1/4}$ remains unchanged and ${\gamma}_{{\mu}{\nu}}$ and ${\cal G}^{{\mu}{\nu}}$
transform as the worldvolume covariant and contravariant tensors, respectively,
which keeps ${\gamma}_{{\mu}{\nu}}{\cal G}^{{\mu}{\nu}}$ invariant.

$S^{{\rm II}}_{\bar{1}}$, different from $S^{{\rm II}}_{1}$ of eq.~(7), is Weyl non-invariant in any dimension of worldvolume because of the non-zero
term of $p+5$ (for further details, see section 4). However, this invariance can be recovered by
the above procedure done for $S^{{\rm II}}_{1}$. That is, introducing an auxiliary scalar field
${\Phi}({\xi})$, and rescaling the worldvolume metric as ${\Phi}{\gamma}_{{\mu}{\nu}}$ in
$S^{{\rm I}}_{\rm P{\bar 1}}$ of eq.~(13), we write down one more parent action associated with the second proposal of construction of parent actions,
\begin{equation}
S^{{\rm I}}_{{\rm P}{\bar{2}}}=-T_{p}\int d^{p+1}{\xi}e^{-\phi}\left[{\Phi}^{(p+1)/4}(-g)^{1/4}(-{\gamma})^{1/4}
+{\Lambda}_{{\mu}{\nu}}\left({\Phi}^{-1}{\gamma}^{{\mu}{\nu}}-{\cal G}^{{\mu}{\nu}}\right)\right].
\end{equation}
Simply following the procedure under eq.~(8), i.e. $S^{{\rm I}}_{{\rm P}{2}}$, we thus deduce another new action with 2-metrics
\begin{equation}
S^{{\rm II}}_{\bar{2}}=-\frac{T_{p}}{4}\int d^{p+1}{\xi}e^{-\phi}(-g)^{1/4}(-{\gamma})^{1/4}\left[-{\Phi}^{(p+5)/4}{\gamma}_{{\mu}{\nu}}
{\cal G}^{{\mu}{\nu}}+(p+5){\Phi}^{(p+1)/4}
\right],
\end{equation}
which restores the Weyl invariance. Furthermore, the other Weyl invariant form can be derived by eliminating the auxiliary scalar field
${\Phi}({\xi})$ from $S^{{\rm II}}_{\bar{2}}$ as the way that $S^{{\rm II}}_{3}$ has been deduced from $S^{{\rm II}}_{2}$,
\begin{equation}
S^{{\rm II}}_{\bar{3}}=-T_{p}\int d^{p+1}{\xi}e^{-\phi}(-g)^{1/4}(-{\gamma})^{1/4}\left(\frac{1}{p+1}{\gamma}_{{\mu}{\nu}}{\cal G}^{{\mu}{\nu}}
\right)^{-(p+1)/4}.
\end{equation}
Obviously $S^{{\rm II}}_{\bar{i}}$ (${\bar{i}}={\bar{1}},{\bar{2}},{\bar{3}}$) contain highly nonlinear terms of field strengths because of our
special definition of ${\cal G}^{{\mu}{\nu}}$.

Similar to the situation of $S^{{\rm II}}_{i}$ ($i=1,2,3$), we can also write down other parent actions besides $S^{{\rm I}}_{{\rm P}{\bar{1}}}$
and $S^{{\rm I}}_{{\rm P}{\bar{2}}}$, however, we obtain no more new forms of actions with 2-metrics under our second proposal but more
dualities among the three new forms $S^{{\rm II}}_{\bar{i}}$ (${\bar{i}}={\bar{1}},{\bar{2}},{\bar{3}}$) that, together with the 1-metric source
action, also constitute a closed set of dual actions. The schematic representation of dual sets, combined with that of $S^{{\rm II}}_{i}$ ($i=1,2,3$),
is shown in Figure 1.

As a consequence, if we apply the parent action method to the 1-metric form with respect to ${\cal G}_{{\mu}{\nu}}$, i.e.,
moving ${\cal G}_{{\mu}{\nu}}$ out of one-fourth root while keeping $g_{{\mu}{\nu}}$ unchanged, we obtain the six 2-metric actions:
$S^{{\rm II}}_{i}$ and $S^{{\rm II}}_{\bar{i}}$, where $i=1,2,3$ and ${\bar{i}}={\bar{1}},{\bar{2}},{\bar{3}}$.
The first three forms with subscript $i$ are derived in terms of the parent action method by adding a Lagrange multiplier term
that associates with ${\Lambda}^{{\mu}{\nu}}$ (a
contravariant Lagrange multiplier tensor field) and ${\cal G}_{{\mu}{\nu}}$,
and have been proposed~\cite{s14,s15} by quite different consideration from ours;
while the other three with subscript ${\bar i}$ are deduced by adding a Lagrange multiplier term
that involves in ${\Lambda}_{{\mu}{\nu}}$ (a covariant Lagrange multiplier tensor field) and the inverse of ${\cal G}_{{\mu}{\nu}}$, i.e.
${\cal G}^{{\mu}{\nu}}$, and are new actions
that contain highly nonlinear terms of field strengths. The two proposals of construction of parent actions divide the six 2-metric forms into two parts,
each of which, together with the 1-metric form, consists of a closed set. Here we mention in advance that this notation will be utilized again in series
{\rm III}, that is, subscript $i$ corresponds to the first proposal of writing parent actions, while subscript ${\bar i}$ corresponds to the
second.

Schematic representation of dual sets composed of the above six actions in series {\rm II} together with series {\rm I} as the source, that is,
($S^{{\rm I}}$, $S^{{\rm II}}_{1}$, $S^{{\rm II}}_{2}$, $S^{{\rm II}}_{3}$) and
($S^{{\rm I}}$, $S^{{\rm II}}_{\bar{1}}$, $S^{{\rm II}}_{\bar{2}}$, $S^{{\rm II}}_{\bar{3}}$), is shown by Figure 1.

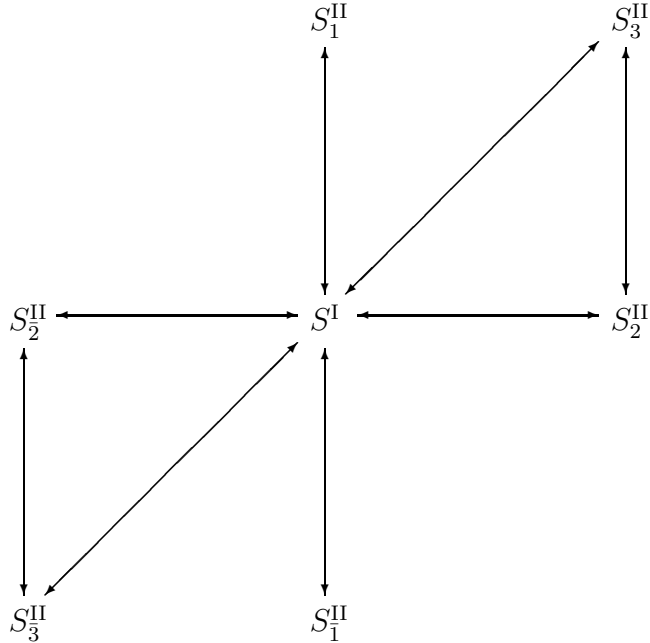
\begin{figure}[htb]
\begin{center}
\setlength{\unitlength}{.8mm}
\begin{picture}(100,100)(0,0)
\put(50,50){$S^{{\rm I}}$}
\put(100,50){$S^{{\rm II}}_{2}$}
\put(58,52.5){\vector(1,0){40}}
\put(98,52.5){\vector(-1,0){40}}
\put(50,100){$S^{{\rm II}}_{1}$}
\put(100,100){$S^{{\rm II}}_{3}$}
\put(52.5,56){\vector(0,1){41}}
\put(52.5,97){\vector(0,-1){41}}
\put(102.5,56){\vector(0,1){41}}
\put(102.5,97){\vector(0,-1){41}}
\put(98,98){\vector(-1,-1){42}}
\put(56,56){\vector(1,1){42}}
\put(0,50){$S^{{\rm II}}_{\bar{2}}$}
\put(0,0){$S^{{\rm II}}_{\bar{3}}$}
\put(8,52.5){\vector(1,0){40}}
\put(48,52.5){\vector(-1,0){40}}
\put(50,0){$S^{{\rm II}}_{\bar{1}}$}
\put(52.5,47){\vector(0,-1){41}}
\put(52.5,6){\vector(0,1){41}}
\put(2.5,6){\vector(0,1){41}}
\put(2.5,47){\vector(0,-1){41}}
\put(6,6){\vector(1,1){42}}
\put(48,48){\vector(-1,-1){42}}
\end{picture}
\caption{\small Dualities in the two sets of dual actions,
($S^{{\rm I}}$, $S^{{\rm II}}_{1}$, $S^{{\rm II}}_{2}$, $S^{{\rm II}}_{3}$) and ($S^{{\rm I}}$, $S^{{\rm II}}_{\bar{1}}$, $S^{{\rm II}}_{\bar{2}}$,
$S^{{\rm II}}_{\bar{3}}$). The line with two arrows connects two actions that are dual to each other.
As a source action, the 1-metric form, $S^{{\rm I}}$,
appears in both sets. It is quite noticeable that in each set
the Weyl non-invariant form,
$S^{{\rm II}}_{1}$ ($S^{{\rm II}}_{\bar{1}}$), has no {\em direct}
dualities to its corresponding Weyl invariant forms, $S^{{\rm II}}_{2}$ and $S^{{\rm II}}_{3}$ ($S^{{\rm II}}_{\bar{2}}$
and $S^{{\rm II}}_{\bar{3}}$).}
\end{center}
\end{figure}

The permutation symmetry of series {\rm I}, i.e. $S^{{\rm I}}$ between ${\cal G}_{{\mu}{\nu}}$ and $g_{{\mu}{\nu}}$ makes us consider
a further application of the parent action method with respect to $g_{{\mu}{\nu}}$. This idea is quite natural after our fulfilment to
${\cal G}_{{\mu}{\nu}}$. That is, if we apply the parent action method to the 1-metric action with respect to the induced metric
$g_{{\mu}{\nu}}$ instead, or in other words,
moving $g_{{\mu}{\nu}}$ out of one-fourth root while keeping ${\cal G}_{{\mu}{\nu}}$ unchanged, we then derive six more
2-metric forms of $Dp$-brane actions, denoted by ${S^{\prime}}^{{\rm II}}$ (subscripts suppressed), by completely following the above procedure
in this section,
\begin{equation}
{S^{\prime}}^{{\rm II}}_{1}=-\frac{T_{p}}{4}\int d^{p+1}{\xi}e^{-\phi}{(-\cal G)}^{1/4}(-{\gamma})^{1/4}\left[{\gamma}^{{\mu}{\nu}}g_{{\mu}{\nu}}-(p-3)
\right],
\end{equation}
\begin{equation}
{S^{\prime}}^{{\rm II}}_{2}=-\frac{T_{p}}{4}\int d^{p+1}{\xi}e^{-\phi}{(-\cal G)}^{1/4}(-{\gamma})^{1/4}\left[{\Phi}^{(p-3)/4}{\gamma}^{{\mu}{\nu}}
g_{{\mu}{\nu}}-(p-3){\Phi}^{(p+1)/4}
\right],
\end{equation}
\begin{equation}
{S^{\prime}}^{{\rm II}}_{3}=-T_{p}\int d^{p+1}{\xi}e^{-\phi}{(-\cal G)}^{1/4}(-{\gamma})^{1/4}\left(\frac{1}{p+1}{\gamma}^{{\mu}{\nu}}g_{{\mu}{\nu}}
\right)^{(p+1)/4},
\end{equation}
\begin{equation}
{S^{\prime}}^{{\rm II}}_{\bar{1}}=-\frac{T_{p}}{4}\int d^{p+1}{\xi}e^{-\phi}{(-\cal G)}^{1/4}(-{\gamma})^{1/4}\left[{-\gamma}_{{\mu}{\nu}}g^{{\mu}{\nu}}
+(p+5)\right],
\end{equation}
\begin{equation}
{S^{\prime}}^{{\rm II}}_{\bar{2}}=-\frac{T_{p}}{4}\int d^{p+1}{\xi}e^{-\phi}{(-\cal G)}^{1/4}(-{\gamma})^{1/4}\left[-{\Phi}^{(p+5)/4}{\gamma}_{{\mu}{\nu}}
g^{{\mu}{\nu}}+(p+5){\Phi}^{(p+1)/4}
\right],
\end{equation}
\begin{equation}
{S^{\prime}}^{{\rm II}}_{\bar{3}}=-T_{p}\int d^{p+1}{\xi}e^{-\phi}{(-\cal G)}^{1/4}(-{\gamma})^{1/4}\left(\frac{1}{p+1}{\gamma}_{{\mu}{\nu}}g^{{\mu}{\nu}}
\right)^{-(p+1)/4},
\end{equation}
which appear, to our knowledge, for the first time. They are classically equivalent and,
together with the 1-metric form, constitute the two dual sets of actions,
($S^{{\rm I}}$, ${S^{\prime}}^{{\rm II}}_{1}$, ${S^{\prime}}^{{\rm II}}_{2}$, ${S^{\prime}}^{{\rm II}}_{3}$) and
($S^{{\rm I}}$, ${S^{\prime}}^{{\rm II}}_{\bar{1}}$, ${S^{\prime}}^{{\rm II}}_{\bar{2}}$, ${S^{\prime}}^{{\rm II}}_{\bar{3}}$). Of course,
this kind of separation
or classification depends on the two different ways of construction of parent actions as mentioned above. We note that ${S^{\prime}}^{{\rm II}}$
can be obtained from $S^{{\rm II}}$ simply under the permutation between ${\cal G}_{{\mu}{\nu}}$ and $g_{{\mu}{\nu}}$
(for further details on permutation, see section 4).
Although they do not supply a simpler
formulation related to field strengths, ${S^{\prime}}^{{\rm II}}$ provide us the possibility that richer contexts of new forms beyond series {\rm II}
should be considered. That is the content of our next section.
For the schematic representation of dualities of ${S^{\prime}}^{{\rm II}}$, see Figure 2, which has the same structure as that of Figure 1
only with the replacement of $S^{{\rm II}}$ by ${S^{\prime}}^{{\rm II}}$.

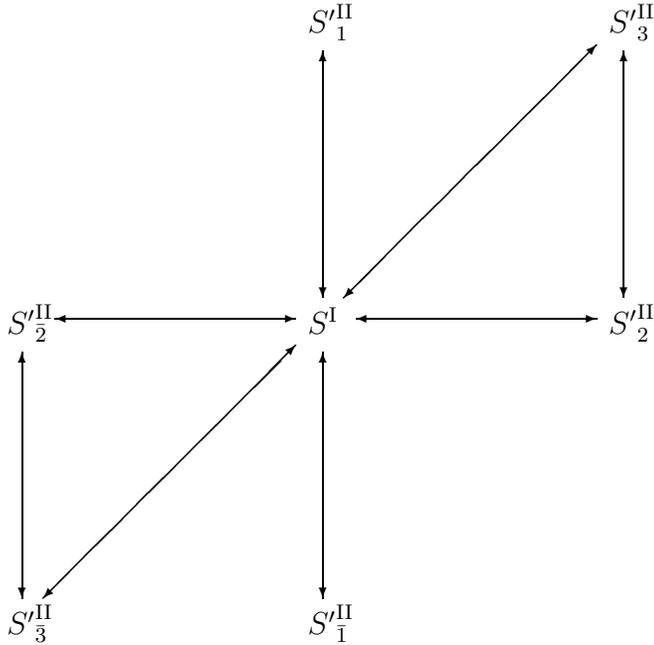
\begin{figure}[htb]
\begin{center}
\setlength{\unitlength}{.8mm}
\begin{picture}(100,100)(0,0)
\put(50,50){$S^{{\rm I}}$}
\put(100,50){${S^{\prime}}^{{\rm II}}_{2}$}
\put(58,52.5){\vector(1,0){40}}
\put(98,52.5){\vector(-1,0){40}}
\put(50,100){${S^{\prime}}^{{\rm II}}_{1}$}
\put(100,100){${S^{\prime}}^{{\rm II}}_{3}$}
\put(52.5,56){\vector(0,1){41}}
\put(52.5,97){\vector(0,-1){41}}
\put(102.5,56){\vector(0,1){41}}
\put(102.5,97){\vector(0,-1){41}}
\put(98,98){\vector(-1,-1){42}}
\put(56,56){\vector(1,1){42}}
\put(0,50){${S^{\prime}}^{{\rm II}}_{\bar{2}}$}
\put(0,0){${S^{\prime}}^{{\rm II}}_{\bar{3}}$}
\put(8,52.5){\vector(1,0){40}}
\put(48,52.5){\vector(-1,0){40}}
\put(50,0){${S^{\prime}}^{{\rm II}}_{\bar{1}}$}
\put(52.5,47){\vector(0,-1){41}}
\put(52.5,6){\vector(0,1){41}}
\put(2.5,6){\vector(0,1){41}}
\put(2.5,47){\vector(0,-1){41}}
\put(6,6){\vector(1,1){42}}
\put(48,48){\vector(-1,-1){42}}
\end{picture}
\caption{\small Dualities in the two sets of dual actions,
($S^{{\rm I}}$, ${S^{\prime}}^{{\rm II}}_{1}$, ${S^{\prime}}^{{\rm II}}_{2}$, ${S^{\prime}}^{{\rm II}}_{3}$) and ($S^{{\rm I}}$,
${S^{\prime}}^{{\rm II}}_{\bar{1}}$, ${S^{\prime}}^{{\rm II}}_{\bar{2}}$,
${S^{\prime}}^{{\rm II}}_{\bar{3}}$). The line with two arrows connects two actions that are dual to each other.
As a source action, the 1-metric form, $S^{{\rm I}}$, appears in both sets. It is quite noticeable that in each set
the Weyl non-invariant form,
${S^{\prime}}^{{\rm II}}_{1}$ (${S^{\prime}}^{{\rm II}}_{\bar{1}}$), has no {\em direct}
dualities to its corresponding Weyl invariant forms, ${S^{\prime}}^{{\rm II}}_{2}$ and ${S^{\prime}}^{{\rm II}}_{3}$
(${S^{\prime}}^{{\rm II}}_{\bar{2}}$ and ${S^{\prime}}^{{\rm II}}_{\bar{3}}$).}
\end{center}
\end{figure}

As a summary,
in series {\rm II}, i.e., 2-metric series, which contains one induced metric $g_{{\mu}{\nu}}$ and
one auxiliary worldvolume (sometimes called {\em intrinsic}) metric ${\gamma}_{{\mu}{\nu}}$ (both of them are symmetric),
we obtain 12 different forms of actions that are equivalent at classical level.
As of derivation of series {\rm II}, six of them, $S^{{\rm II}}_{i}$ and $S^{{\rm II}}_{\bar{i}}$, where $i=1,2,3$ and
${\bar{i}}={\bar{1}},{\bar{2}},{\bar{3}}$, are related to the application of the parent action method to series {\rm I} with respect to
${\cal G}_{{\mu}{\nu}}$, while the other six, ${S^{\prime}}^{{\rm II}}_{i}$ and ${S^{\prime}}^{{\rm II}}_{\bar{i}}$, however, with respect to
$g_{{\mu}{\nu}}$. Moreover, in both of $S^{{\rm II}}$ and ${S^{\prime}}^{{\rm II}}$, the two proposals for construction of parent actions
are considered. The clear advantage of this method exists, we may say, in the two aspects, one is its systematicness and the other its natural
connection with duality as we have seen in the above discussions. The result shows a richer context of new actions than that of bosonic $p$-branes
in which only appears the induced metric
$g_{{\mu}{\nu}}$ in the Nambu-Goto form as a source action~\cite{s26}.

\section{3-Metric Form and Duality}

The symmetric status of ${\cal G}_{{\mu}{\nu}}$ and $g_{{\mu}{\nu}}$ in series {\rm I}, i.e. $S^{{\rm I}}$ as the original form
provides us the possibility to go further beyond the region of series {\rm II}.
That is, we can consider a new kind of action forms that are {\em not only} quadratic in the abelian field strength {\em but also}
``formally quadratic'' in the coordinate of spacetime. In other words, the new actions are {\em not only} linear in ${\cal G}_{{\mu}{\nu}}$
{\em but also} seemingly linear in $g_{{\mu}{\nu}}$.
Because the definition of ${\cal G}_{{\mu}{\nu}}$ contains the inverse of the induced metric $g_{{\mu}{\nu}}$, the new forms are not able to be quadratic
in the coordinate of spacetime but {\em gain an advantage over one-fourth roots} that are non-rational.
If we try to move both ${\cal G}_{{\mu}{\nu}}$ and $g_{{\mu}{\nu}}$ out of one-fourth roots by following the
same procedure as that from series {\rm I} to
series {\rm II}, one direct way is to introduce one more auxiliary worldvolume tensor field ${\omega}_{{\mu}{\nu}}$
that will play a similar role to that of
the intrinsic metric ${\gamma}_{{\mu}{\nu}}$ introduced in series {\rm II}.  For a deeper understanding of the relation between
${\gamma}_{{\mu}{\nu}}$ and ${\omega}_{{\mu}{\nu}}$, see the next section.
If so, there will appear three metrics in an action all of which are symmetric, that is, one induced, one intrinsic and
one newly introduced intrinsic-like metrics.
According to the result in section 2,
six 3-metric forms will be deduced from each 2-metric source action in terms of the parent action method together with the two proposals of constructing
parent actions. We thus have $6\times6=36$
new forms of 3-metric actions, denoted by $S^{{\rm III}}$ (subscripts suppressed here and explained in the following paragraphs),
by treating the six 2-metric forms, $S^{{\rm II}}_{i}$ and
$S^{{\rm II}}_{\bar{i}}$,
as source actions;
and moreover, we still have the other 36 new forms, ${S^{\prime}}^{{\rm III}}$ (subscripts suppressed here and explained in the following paragraphs),
by using the remaining six 2-metric source forms,
${S^{\prime}}^{{\rm II}}_{i}$ and
${S^{\prime}}^{{\rm II}}_{\bar{i}}$, where
${i}=1,2,3$ and ${\bar{i}}={\bar{1}},{\bar{2}},{\bar{3}}$.
Consequently, we will acquire
$36+36=72$
newly proposed 3-metric
forms in total.

In series {\rm III}, i.e., 3-metric series, which includes an additional worldvolume metric ${\omega}_{{\mu}{\nu}}$
besides the induced and intrinsic ones,
72 new forms of $Dp$-brane actions can be derived as we stated above. Alternatively, they can also be obtained by treating the 1-metric action
$S^{{\rm I}}$ as the
source and doing two successive times of application of the parent action method to it.
If we apply the parent action method to the 1-metric action with respect firstly to
${\cal G}_{{\mu}{\nu}}$, and secondly to the induced metric $g_{{\mu}{\nu}}$, called order (${\cal G}_{{\mu}{\nu}}$, $g_{{\mu}{\nu}}$), that is,
moving
${\cal G}_{{\mu}{\nu}}$ out of its one-fourth root at first and then moving $g_{{\mu}{\nu}}$,
we therefore gain the 36 forms $S^{{\rm III}}$. As the derivation of the new 3-metric forms is, though tedious, straightforward by following
one of the two schemes (one is to apply the parent action method just once by treating the 2-metric forms as source actions, and the other is to
apply the parent action method twice but starting from the 1-metric form as the source), we omit
the procedure but just simply list
the forms as a whole in Appendix A. However, we write down, as an example, one of typical actions with 3-metrics as follows:
\begin{equation}
S^{{\rm III}}_{11}=-\frac{T_{p}}{16}\int d^{p+1}{\xi}e^{-\phi}(-{\gamma})^{1/4}(-{\omega})^{1/4}\Big[{\gamma}^{{\mu}{\nu}}{\cal G}_{{\mu}{\nu}}-(p-3)
\Big]\Big[{\omega}^{{\lambda}{\sigma}}g_{{\lambda}{\sigma}}-(p-3)\Big],
\end{equation}
which is obtained by following the order (${\cal G}_{{\mu}{\nu}}$, $g_{{\mu}{\nu}}$) with the consideration of the first proposal of constructing parent
actions for both ${\cal G}_{{\mu}{\nu}}$ and $g_{{\mu}{\nu}}$. This is a rational form in the induced metric, which gains an advantage over one-fourth
roots with the price paid by introducing one more auxiliary worldvolume metric ${\omega}_{{\mu}{\nu}}$ as was mentioned in the above paragraph.
When describing $D3$-branes, i.e. eq.~(33), this action form turns to possess the double Weyl invariance (see the next section for details).
Moreover, if we do in the reverse order, i.e., order
($g_{{\mu}{\nu}}$, ${\cal G}_{{\mu}{\nu}}$), we get the other 36 forms ${S^{\prime}}^{{\rm III}}$ that are listed in Appendix B.
Of course, the 72 forms are
equivalent at classical level, that is, all of them give rise to the same equations of motion.
Note that in the notation of series {\rm III} two subscripts are needed in which subscript $i$ corresponds to the first proposal of
writing parent actions, while subscript
${\bar i}$ corresponds to the second because of twice of application of the parent action method. This is different from the case of series {\rm II}
in which the parent action method is applied only once.
For example, the symbol $S^{\rm III}_{i{\bar j}}$ denotes (a) $S^{\rm III}$ corresponds to order (${\cal G}_{{\mu}{\nu}}$, $g_{{\mu}{\nu}}$),
and (b) subscript $i$ corresponds to firstly moving ${\cal G}_{{\mu}{\nu}}$ in terms of the first proposal
while subscript ${\bar j}$ to secondly moving $g_{{\mu}{\nu}}$
in terms of the second proposal. The other symbols, such as $S^{\rm III}_{ij}$, $S^{\rm III}_{{\bar i}j}$, and $S^{\rm III}_{{\bar i}{\bar j}}$, depend
on the other modes of permutations and combinations of twice use of the first and second proposals.
In addition, we emphasize that one more auxiliary scalar field
$\Psi(\xi)$ is necessary
to be introduced for restoration of single Weyl invariance or for further improvement to double Weyl invariance
besides $\Phi(\xi)$ for series {\rm III} because of the existence
of two intrinsic metrics,
where $\Phi(\xi)$ and $\Psi(\xi)$, functions of worldvolume parameters, are scalars
in both the spacetime and worldvolume, and correspond to ${\gamma}_{{\mu}{\nu}}$ and ${\omega}_{{\mu}{\nu}}$, respectively. For details of
the single and double Weyl invariance, see the next section.

It is much more complicated to classify series {\rm III} into dual sets than to classify series {\rm II} because it depends on two successive
times of use of the
parent action method with respect to ${\cal G}_{{\mu}{\nu}}$ and
$g_{{\mu}{\nu}}$, respectively, and on the consideration of two proposals of construction of parent actions.
There are different kinds of classification for different choices of order of
${\cal G}_{{\mu}{\nu}}$ and
$g_{{\mu}{\nu}}$ and of order of the two proposals. However, one direct and convenient approach is to adopt the classification for series {\rm II},
which results, of course, in more number of dual sets in series {\rm III}.
In this way, 36 $Dp$-brane actions listed in Appendix A, i.e. $S^{{\rm III}}$ can be classified into 12 dual sets.
Schematic representations of the 12 dual sets between each 2-metric source action and its corresponding six 3-metric forms are shown in Figures 3 \& 4.
Similarly, ${S^{\prime}}^{{\rm III}}$ constitute their own 12 dual sets of actions
if we adopt the same classification as that used for $S^{{\rm III}}$. We ignore the schematic representations of dual sets
for ${S^{\prime}}^{{\rm III}}$ but just
mention that they have the completely same structure as that of Figures 3 \& 4 with only the replacement of all $S$ (including both series {\rm II} and
series {\rm III}) by $S^{\prime}$.

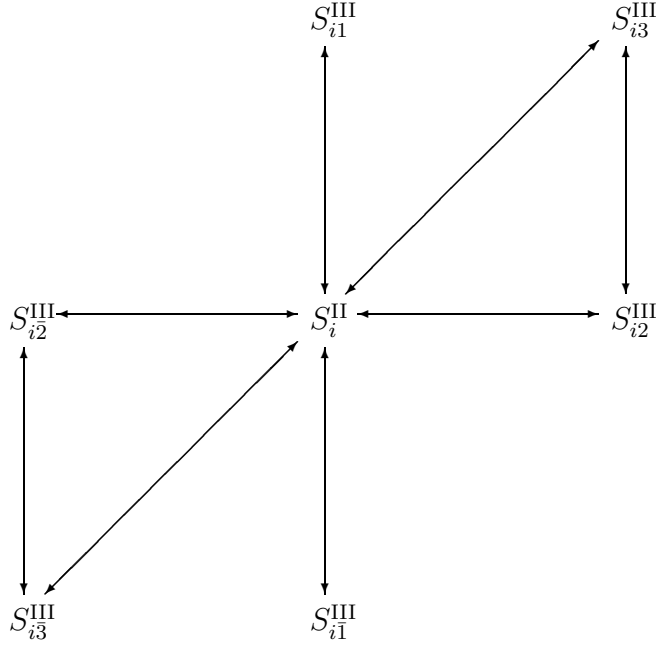
\begin{figure}[htb]
\begin{center}
\setlength{\unitlength}{.8mm}
\begin{picture}(100,100)(0,0)
\put(50,50){$S^{{\rm II}}_{i}$}
\put(100,50){$S^{{\rm III}}_{i2}$}
\put(58,52.5){\vector(1,0){40}}
\put(98,52.5){\vector(-1,0){40}}
\put(50,100){$S^{{\rm III}}_{i1}$}
\put(100,100){$S^{{\rm III}}_{i3}$}
\put(52.5,56){\vector(0,1){41}}
\put(52.5,97){\vector(0,-1){41}}
\put(102.5,56){\vector(0,1){41}}
\put(102.5,97){\vector(0,-1){41}}
\put(98,98){\vector(-1,-1){42}}
\put(56,56){\vector(1,1){42}}
\put(0,50){$S^{{\rm III}}_{i{\bar 2}}$}
\put(0,0){$S^{{\rm III}}_{i{\bar 3}}$}
\put(8,52.5){\vector(1,0){40}}
\put(48,52.5){\vector(-1,0){40}}
\put(50,0){$S^{{\rm III}}_{i{\bar 1}}$}
\put(52.5,47){\vector(0,-1){41}}
\put(52.5,6){\vector(0,1){41}}
\put(2.5,6){\vector(0,1){41}}
\put(2.5,47){\vector(0,-1){41}}
\put(6,6){\vector(1,1){42}}
\put(48,48){\vector(-1,-1){42}}
\end{picture}
\caption{\small Dualities in the six sets of dual actions,
($S^{{\rm II}}_{i}$, $S^{{\rm III}}_{i1}$, $S^{{\rm III}}_{i2}$, $S^{{\rm III}}_{i3}$) and ($S^{{\rm II}}_{i}$, $S^{{\rm III}}_{i{\bar 1}}$,
$S^{{\rm III}}_{i{\bar 2}}$, $S^{{\rm III}}_{i{\bar 3}}$), where $i=1,2,3$.
The line with two arrows connects two actions that are dual to each other.
As a source action, each 2-metric form, $S^{{\rm II}}_{i}$, appears in its corresponding two sets. It is quite noticeable that in each set
the Weyl non-invariant form,
$S^{{\rm III}}_{i1}$ ($S^{{\rm III}}_{i{\bar 1}}$), has no {\em direct}
dualities to its corresponding Weyl invariant forms, $S^{{\rm III}}_{i2}$ and $S^{{\rm III}}_{i3}$ ($S^{{\rm III}}_{i{\bar 2}}$
and $S^{{\rm III}}_{i{\bar 3}}$).}
\end{center}
\end{figure}

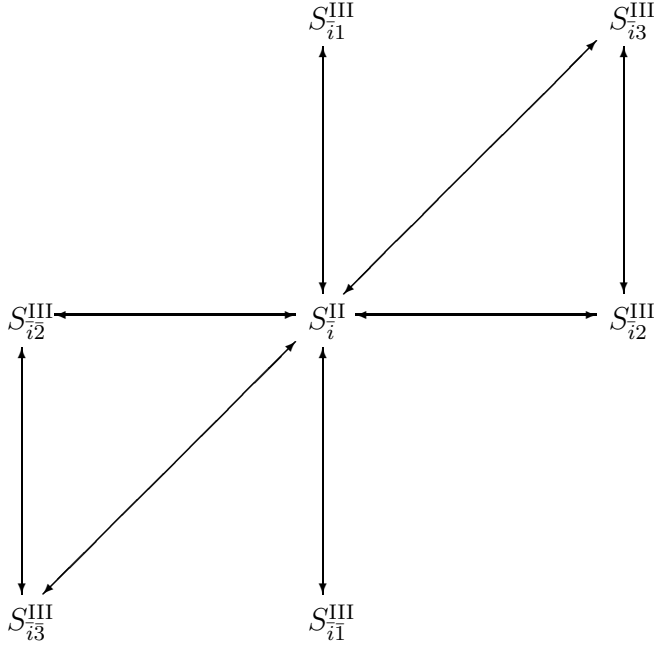
\begin{figure}[htb]
\begin{center}
\setlength{\unitlength}{.8mm}
\begin{picture}(100,100)(0,0)
\put(50,50){$S^{{\rm II}}_{{\bar i}}$}
\put(100,50){$S^{{\rm III}}_{{\bar i}2}$}
\put(58,52.5){\vector(1,0){40}}
\put(98,52.5){\vector(-1,0){40}}
\put(50,100){$S^{{\rm III}}_{{\bar i}1}$}
\put(100,100){$S^{{\rm III}}_{{\bar i}3}$}
\put(52.5,56){\vector(0,1){41}}
\put(52.5,97){\vector(0,-1){41}}
\put(102.5,56){\vector(0,1){41}}
\put(102.5,97){\vector(0,-1){41}}
\put(98,98){\vector(-1,-1){42}}
\put(56,56){\vector(1,1){42}}
\put(0,50){$S^{{\rm III}}_{{\bar i}{\bar 2}}$}
\put(0,0){$S^{{\rm III}}_{{\bar i}{\bar 3}}$}
\put(8,52.5){\vector(1,0){40}}
\put(48,52.5){\vector(-1,0){40}}
\put(50,0){$S^{{\rm III}}_{{\bar i}{\bar 1}}$}
\put(52.5,47){\vector(0,-1){41}}
\put(52.5,6){\vector(0,1){41}}
\put(2.5,6){\vector(0,1){41}}
\put(2.5,47){\vector(0,-1){41}}
\put(6,6){\vector(1,1){42}}
\put(48,48){\vector(-1,-1){42}}
\end{picture}
\caption{\small Dualities in the six sets of dual actions,
($S^{{\rm II}}_{{\bar i}}$, $S^{{\rm III}}_{{\bar i}1}$, $S^{{\rm III}}_{{\bar i}2}$, $S^{{\rm III}}_{{\bar i}3}$) and ($S^{{\rm II}}_{{\bar i}}$,
$S^{{\rm III}}_{{\bar i}{\bar 1}}$,
$S^{{\rm III}}_{{\bar i}{\bar 2}}$, $S^{{\rm III}}_{{\bar i}{\bar 3}}$), where ${\bar i}={\bar 1},{\bar 2},{\bar 3}$.
The line with two arrows connects two actions that are dual to each other.
As a source action, each 2-metric form, $S^{{\rm II}}_{{\bar i}}$, appears in its corresponding two sets. It is quite noticeable that in each set
the Weyl non-invariant form,
$S^{{\rm III}}_{{\bar i}1}$ ($S^{{\rm III}}_{{\bar i}{\bar 1}}$), has no {\em direct}
dualities to its corresponding Weyl invariant forms, $S^{{\rm III}}_{{\bar i}2}$ and $S^{{\rm III}}_{{\bar i}3}$ ($S^{{\rm III}}_{{\bar i}{\bar 2}}$
and $S^{{\rm III}}_{{\bar i}{\bar 3}}$).}
\end{center}
\end{figure}

\section{Permutation and Weyl Symmetries}

Nonetheless, the 72 forms of series {\rm III} have some relations among themselves that might be interesting. If we divide them by the kinds of
orders (${\cal G}_{{\mu}{\nu}}$, $g_{{\mu}{\nu}}$) and ($g_{{\mu}{\nu}}$, ${\cal G}_{{\mu}{\nu}}$) that are related to the ways of
following the parent action method and of proposing
possible parent actions, we then obtain two parts or groups, $S^{\rm III}$ and ${S^{\prime}}^{{\rm III}}$, each of which contains 36 actions as listed
in Appendix A and Appendix B.
We find that $S^{\rm III}$ changes to ${S^{\prime}}^{{\rm III}}$ and {\em vice versa} under the permutation transformation between the intrinsic metric
${\gamma}_{{\mu}{\nu}}$ and the intrinsic-like metric
${\omega}_{{\mu}{\nu}}$, and simultaneously between their corresponding auxiliary scalar fields $\Phi({\xi})$ and $\Psi({\xi})$, that is,
\begin{eqnarray}
{\gamma}_{{\mu}{\nu}} \rightleftharpoons {\omega}_{{\mu}{\nu}}, \hspace{1.5cm}
\Phi({\xi}) \rightleftharpoons \Psi({\xi}).
\end{eqnarray}
The one-to-one correspondence between the two groups exists as follows:
\begin{eqnarray}
S^{\rm III}_{ij} \rightleftharpoons {S^{\prime}}^{{\rm III}}_{ji}, \hspace{1.5cm}
S^{\rm III}_{i{\bar j}} \rightleftharpoons {S^{\prime}}^{{\rm III}}_{{\bar j}i},\nonumber \\
S^{\rm III}_{{\bar i}j} \rightleftharpoons {S^{\prime}}^{{\rm III}}_{j{\bar i}}, \hspace{1.5cm}
S^{\rm III}_{{\bar i}{\bar j}} \rightleftharpoons {S^{\prime}}^{{\rm III}}_{{\bar j}{\bar i}},
\end{eqnarray}
where $i,j=1,2,3$ and ${\bar i},{\bar j}={\bar 1},{\bar 2},{\bar 3}$. Such a correspondence shows that ${\gamma}_{{\mu}{\nu}}$ and
${\omega}_{{\mu}{\nu}}$ have an equivalent status in series {\rm III}. It is not surprising to have the conclusion because both of them are auxiliary
worldvolume tensor fields. This is the reason that we call ${\omega}_{{\mu}{\nu}}$ an intrinsic-like metric.
As to series {\rm II}, although a similar one-to-one correspondence,
$S^{\rm II}_{i} \rightleftharpoons {S^{\prime}}^{{\rm II}}_{i}$ and $S^{\rm II}_{{\bar i}} \rightleftharpoons {S^{\prime}}^{{\rm II}}_{{{\bar i}}}$,
appears under the permutation between ${\cal G}_{{\mu}{\nu}}$ and $g_{{\mu}{\nu}}$, this just reflects our application of the parent
action method to series {\rm I} with respect to ${\cal G}_{{\mu}{\nu}}$ or $g_{{\mu}{\nu}}$ but nothing else.

Let us turn to investigate Weyl invariance of $Dp$-brane actions in series {\rm II} and series {\rm III}. We just consider $S^{\rm II}$ and
$S^{\rm III}$. As to ${S^{\prime}}^{{\rm II}}$ and ${S^{\prime}}^{{\rm III}}$, similar results can be
acquired through the permutation between $S$ and ${S^{\prime}}$ (series {\rm II} and series {\rm III}, respectively).

In the six forms of $S^{\rm II}$, $S^{\rm II}_{1}$ is Weyl non-invariant for the general case of $p \neq 3$, and $S^{\rm II}_{{\bar 1}}$ that contains
highly nonlinear terms of field strengths is Weyl non-invariant in any dimension of worldvolume. The Weyl non-invariance of $S^{\rm II}_{{\bar 1}}$
is {\em inevitable} because of the definition of the inverse formulation ${\cal G}^{{\mu}{\nu}}$ which brings about the non-zero term of $p+5$,
and a similar phenomenon happened
in one of new forms for string theory~\cite{s26}. However, the remaining four forms,
\begin{equation}
\begin{array}{cccc}
S^{\rm II}_{2},& S^{\rm II}_{3};& S^{\rm II}_{{\bar 2}},& S^{\rm II}_{{\bar 3}},
\end{array}
\end{equation}
as they have been proposed, keep Weyl invariant for the general case under the conformal transformation:
\begin{eqnarray}
{\gamma}_{{\mu}{\nu}}({\xi}) & \longrightarrow & {\rm exp}\left({\eta}({\xi})\right)
{\gamma}_{{\mu}{\nu}}({\xi}), \nonumber \\
{\Phi}({\xi}) & \longrightarrow & {\rm exp}\left(-{\eta}({\xi})\right)
{\Phi}({\xi}),
\label{wt1}
\end{eqnarray}
where ${\eta}({\xi})$ is an arbitrary real function of worldvolume parameters. For the special case of $p=3$, $S^{\rm II}_{1}$ of $D3$-branes
restores the Weyl
invariance, which is similar to the string case of bosonic $p$-branes~\cite{s23}. Moreover, $S^{\rm II}_{2}$, $S^{\rm II}_{3}$, $S^{\rm II}_{{\bar 2}}$,
and
$S^{\rm II}_{{\bar 3}}$ retain their Weyl invariance, but both $S^{\rm II}_{2}$ and $S^{\rm II}_{3}$ degenerate into $S^{\rm II}_{1}$
in this special case. We note that this kind of degeneracy also appears in series {\rm III} but, however, gives rise to richer contents,
see below for details.
Therefore, there are only four independent 2-metric forms in $S^{\rm II}$, i.e., $S^{\rm II}_{1}$ and $S^{\rm II}_{{\bar i}}$
(${\bar i}={\bar 1},{\bar 2},{\bar 3}$), that describe $D3$-branes.

It seems to be more complicated but is in fact more interesting to analyze the Weyl invariance for series {\rm III}
as two intrinsic metrics (${\gamma}_{{\mu}{\nu}}$ and ${\omega}_{{\mu}{\nu}}$) appear in every form of actions.
Here we point out in advance that a new phenomenon that does not exist
in series {\rm II} is that some actions possess a so-called bi-Weyl or double Weyl invariance. It is the appearance of two intrinsic metrics that
the new phenomenon occurs. This shows the richness of
series {\rm III} {\em not only} in number of various forms as mentioned before {\em but also} in the aspect of Weyl invariance if compared with
series {\rm II}.

We now analyze Weyl invariance for the thirty-six forms of $S^{\rm III}$ in the general case of $p \neq 3$.
Four of them,
\begin{equation}
\begin{array}{cccc}
S^{{\rm III}}_{11}, & S^{{\rm III}}_{1{\bar 1}}, & S^{{\rm III}}_{{\bar 1}1}, & S^{\rm III}_{{\bar 1}{\bar 1}},
\end{array}
\end{equation}
do not have such an invariance.
In particular, $S^{\rm III}_{{\bar 1}{\bar 1}}$ is Weyl non-invariant in any dimension of worldvolume, which is {\em inevitable}
because of the non-zero term related to $p+5$ as happened to $S^{{\rm II}}_{{\bar 1}}$ in series {\rm II}.
However, 16 forms of series {\rm III}
\begin{equation}
\left\{
\begin{array}{cccc}
S^{{\rm III}}_{i1},& S^{{\rm III}}_{i{\bar 1}},& S^{{\rm III}}_{{\bar i}1},& S^{{\rm III}}_{{\bar i}{\bar 1}}; \\
S^{{\rm III}}_{1i},& S^{{\rm III}}_{{\bar 1}i},& S^{{\rm III}}_{1{\bar i}},& S^{{\rm III}}_{{\bar 1}{\bar i}},
\end{array}\right.
\end{equation}
where $i=2,3$ and ${\bar i}={\bar 2},{\bar 3}$, possess the usual, or specifically, {\em single} Weyl invariance. To emphasize {\em single} is just to
distinguish it from the so-called {\em double}
Weyl invariance that will be seen soon.
Eight forms on the first line of eq.~(30)
correspond to the transformation eq.~(28), while the other eight on the second line of eq.~(30) maintain Weyl invariance under the
transformation
\begin{eqnarray}
{\omega}_{{\mu}{\nu}}({\xi}) & \longrightarrow & {\rm exp}\left({\rho}({\xi})\right)
{\omega}_{{\mu}{\nu}}({\xi}), \nonumber \\
{\Psi}({\xi}) & \longrightarrow & {\rm exp}\left(-{\rho}({\xi})\right)
{\Psi}({\xi}),
\label{wt2}
\end{eqnarray}
where ${\rho}({\xi})$ is another arbitrary real function of worldvolume parameters that is, in general, different from ${\eta}({\xi})$.
At last, the remaining 16 forms
of $Dp$-brane actions with 3-metrics
\begin{equation}
\begin{array}{cccc}
S^{{\rm III}}_{ij},& S^{{\rm III}}_{i{\bar j}},& S^{{\rm III}}_{{\bar j}i},& S^{{\rm III}}_{{\bar i}{\bar j}},
\end{array}
\end{equation}
where $i,j=2,3$ and ${\bar i},{\bar j}={\bar 2},{\bar 3}$,
possess {\em double} Weyl invariance, that is, they are invariant under both conformal transformations eqs.~(28) and~(31).
Because there are two intrinsic metrics in series {\rm III} (the corresponding scalar fields ${\Phi}({\xi})$ and ${\Psi}({\xi})$ appear
only in some of 3-metric forms),
two Weyl transformations occur naturally.
This is the new symmetry we have discovered to $Dp$-branes.

For the special case of $p=3$, the investigation of Weyl invariance for series {\rm III} is of particular interests.
To the first three forms of eq.~(29), $S^{{\rm III}}_{11}$, reduced to be
\begin{equation}
S^{{\rm III}}_{11}(p=3)=-\frac{T_{3}}{16}\int d^{4}{\xi}e^{-\phi}(-{\gamma})^{1/4}(-{\omega})^{1/4}\Big({\gamma}^{{\mu}{\nu}}{\cal G}_{{\mu}{\nu}}
\Big)\Big({\omega}^{{\lambda}{\sigma}}g_{{\lambda}{\sigma}}\Big),
\end{equation}
where ${\mu}, {\nu}, {\lambda}, {\sigma}=0,1,2,3$, now possesses the double Weyl invariance; $S^{{\rm III}}_{1{\bar 1}}$ and $S^{{\rm III}}_{{\bar 1}1}$
restore the single Weyl invariance
related to the transformations eq.~(28)
and eq.~(31), respectively. In eq.~(30) the eight forms with subscript $1$,
\begin{equation}
\begin{array}{ccccc}
S^{{\rm III}}_{i1},& S^{{\rm III}}_{{\bar i}1},& S^{{\rm III}}_{1i},& S^{{\rm III}}_{1{\bar i}}, & (i=2,3; \hspace{2mm}{\bar i}={\bar 2},{\bar 3}),
\end{array}
\end{equation}
now extend their Weyl invariance from {\em single} to {\em double},
while the other eight with subscript ${\bar 1}$,
\begin{equation}
\begin{array}{ccccc}
S^{{\rm III}}_{i{\bar 1}}, & S^{{\rm III}}_{{\bar i}{\bar 1}}, & S^{{\rm III}}_{{\bar 1}i},& S^{{\rm III}}_{{\bar 1}{\bar i}},
& (i=2,3; \hspace{2mm}{\bar i}={\bar 2},{\bar 3}),
\end{array}
\end{equation}
keep their original single Weyl invariance unchanged. Finally, the 16 forms in eq.~(32) still maintain
the double Weyl invariance in the special case, which is obvious as they have such a new invariance in any dimension of worldvolume.

In this special case it seems that more actions restore single or double Weyl invariance, however, it is, in fact, that some forms of
actions coincide with each other and
thus the number of independent forms of $S^{{\rm III}}$ decreases from 36 to 16. Note that a similar but simpler case occurred to $S^{{\rm II}}$
as mentioned above.
We classify the 16 independent forms into three sets
in accordance with their Weyl invariance. The first set with {\em no} Weyl invariance contains only one action $S^{{\rm III}}_{{\bar 1}{\bar 1}}$ that is
Weyl non-invariant in any dimension of worldvolume as we pointed out before. The second set with {\em single} Weyl invariance includes six
different forms of actions
\begin{equation}
\begin{array}{cccc}
(S^{{\rm III}}_{i{\bar 1}}), & (S^{{\rm III}}_{{\bar 1}i}), & S^{{\rm III}}_{{\bar i}{\bar 1}}, & S^{{\rm III}}_{{\bar 1}{\bar i}},
\end{array}
\end{equation}
where $i=1,2,3$ and ${\bar i}={\bar 2},{\bar 3}$, and a bracket means that action forms inside coincide with each other.
That is, the independent number of actions inside a bracket is one.
The last set with {\em double}
Weyl invariance takes the following nine forms of actions
\begin{equation}
\begin{array}{cccccc}
(S^{{\rm III}}_{ij}), & (S^{{\rm III}}_{i{\bar 2}}), & (S^{{\rm III}}_{i{\bar 3}}), & (S^{{\rm III}}_{{\bar 2}i}), & (S^{{\rm III}}_{{\bar 3}i}),
& S^{{\rm III}}_{{\bar i}{\bar j}},
\end{array}
\end{equation}
where $i,j=1,2,3$ and ${\bar i},{\bar j}={\bar 2},{\bar 3}$, and the bracket has the same meaning as above.
To the notations of eq.~(36) and eq.~(37), we give a brief explanation. The first term of eq.~(37), for instance,
means that $S^{{\rm III}}_{i1}$ and $S^{{\rm III}}_{1i}$ of eq.~(34) and $S^{{\rm III}}_{ij}$ of eq.~(32) with
double Weyl invariance, eight forms in total, are reduced to
$S^{{\rm III}}_{11}$ of eq.~(33). We may utilize ``degree of degeneracy'' to describe this situation happened in the case of $p=3$.
If so, we may say the degree
of degeneracy for $S^{{\rm III}}_{11}$ is 9. Other terms with brackets have a similar meaning but different degrees of degeneracy.
The terms without brackets have the degree of degeneracy 1.
In a sense, $D3$-branes are singular to the other $Dp$-branes whose dimension of worldvolume is not equal to four. In fact, the $D3$-branes
have played a central role in recent studies of $Dp$-brane dynamics and string theory especially for the $AdS/CFT$ correspondence~\cite{s29}.
Among the
sixteen 3-metric
forms of $D3$-branes, $S^{{\rm III}}_{11}$ of eq.~(33) with the largest degree of degeneracy is the simplest and, in particular, double
Weyl invariant.
It is known~\cite{s18,s19} that the Weyl invariance of string theory has greatly simplified the theory's analysis and allowed its covariant quantization.
The {\em double} Weyl invariance of $D3$-branes ($Dp$-branes in general) may shed some light in this direction.

\section{Conclusion}

In fact, this paper focuses on a natural extension of our previous work~\cite{s26} from bosonic $p$-branes to $Dp$-branes, that is,
deriving various forms of
$Dp$-brane actions and establishing their dualities in terms of the parent action method. In ref.~\cite{s26}, we mentioned that the
parent action method could be applied to $Dp$-branes and thought naively that parallel, or precisely speaking, trivial conclusions would be made.
However, our discussions above
are quite non-trivial, which shows much richer contents on new forms, dualities and conformal symmetries of $Dp$-branes than that of $p$-branes.
The richness and/or non-triviality presents, in particular, in the two aspects: (a) both 2-metric and 3-metric prescriptions are considered, and (b)
an interesting new symmetry, i.e. the double conformal invariance is discovered for the first time in some action forms with 3-metrics.

Under the 2-metric prescription, we obtain 12 different forms of $Dp$-brane actions which
are classically equivalent, that is, which give rise to the same equations of motion. The forms can be classified into the four groups or sets:
$S^{\rm II}_{i}$, $S^{\rm II}_{{\bar i}}$ and
${S^{\prime}}^{\rm II}_{i}$, ${S^{\prime}}^{\rm II}_{{\bar i}}$, where $i=1,2,3$ and ${\bar i}={\bar 1},{\bar 2},{\bar 3}$. Each set, together
with the 1-metric source action $S^{\rm I}$, consists of a closed set of dual actions, and is shown visually by Figures 1 and 2.
For the conformal symmetry of $S^{\rm II}$, $S^{\rm II}_{1}$ and $S^{\rm II}_{{\bar 1}}$ are Weyl non-invariant, while the
rest (four actions) Weyl invariant in the general case of $p\neq 3$; however, $S^{\rm II}_{i}$ ($i=1,2,3$) coincide with each other
(called in this paper {\em degeneracy} which appears in series {\rm III} with varieties) and
possess the Weyl invariance, and the other three $S^{\rm II}_{{\bar i}}$ (${\bar i}={\bar 1},{\bar 2},{\bar 3}$) keep their Weyl non-invariance
or invariance unchanged in the special case of $p=3$. As to ${S^{\prime}}^{\rm II}$, similar results appear.
The advantage of
the parent action method is the systematicness, which has been pointed out in the study of bosonic $p$-branes~\cite{s26}. That is,
we obtain a series of results that cover, on the one hand,
the known actions (cf. $S^{\rm II}_{i}$) proposed by others~\cite{s14,s15} and provide, especially on the other hand,
new actions and interesting duality structures combining series {\rm I} with series {\rm II}.
We note that the systematicness has brought about more new actions and richer dualities in series {\rm III}.
Moreover, we emphasize that
our two proposals for writing parent actions play an important role in the procedure of deriving new actions
and building their dualities. The importance exists {\em not only} in the 2-metric prescription
{\em but also} in the 3-metric prescription that is going to be summarized next.

The motivation to introduce the additional auxiliary worldvolume metric ${\omega}_{{\mu}{\nu}}$, or in other words,
to propose the 3-metric prescription, lies in the
construction of such an action form that does not involve in the one fourth root of the induced metric $g_{{\mu}{\nu}}$.
That is, our aim is to acquire such a form that is a rational functional of the induced
metric as the rational formalism would probably be useful for us to analyze and/or covariantly quantize it according to the experience
from the string theory~\cite{s18,s19}.
Under the 3-metric prescription, we work out 72 equivalent $Dp$-brane actions, $S^{\rm III}$ and ${S^{\prime}}^{\rm III}$, each of which contains
36 different forms. Although the classification of series {\rm III} into dual sets is more complicated than that of series {\rm II}, and in particular,
is not unique but depends on different procedures followed by for deriving the actions,
it becomes easier and the resulting duality structure is simpler if we adopt
the approach utilized in series {\rm II}. For $S^{\rm III}$, 36 forms are classified into 12 dual sets as shown in detail by Figures 3 and 4,
which is the most convenient classification. As to ${S^{\prime}}^{\rm III}$, the same duality structure exists.
A quite interesting relation between $S^{\rm III}$ and ${S^{\prime}}^{\rm III}$ is the permutation under the transformation eq.~(25), and the concrete
one-to-one
correspondence of actions is given by eq.~(26). It is the relation that we conclude the equivalent status of the two auxiliary worldvolume metrics
${\gamma}_{{\mu}{\nu}}$ and ${\omega}_{{\mu}{\nu}}$ in series {\rm III}. We note that all the three metrics,  $g_{{\mu}{\nu}}$, ${\gamma}_{{\mu}{\nu}}$
and ${\omega}_{{\mu}{\nu}}$ appear in a symmetric way in each action of series {\rm III}.
In addition, the Weyl non-invariance and invariance
of all action forms of series {\rm III}
have been analyzed completely for the general case of $p\neq 3$ and for the special case of $p=3$ as well in section 4.
Here what we want to emphasize for 3-metric actions is the double
Weyl invariance that the 16 forms in eq.~(32) possess. This symmetry is associated closely with the appearance of the two worldvolume metrics
and pointed out in $Dp$-branes, to our knowledge, for the first time.

The supersymmetric generalization of the action of the Born-Infeld type eq.~(1) or of series {\rm I} is straightforward~\cite{s9,s10,s11,s12}.
This seems to be maintained in some forms of series {\rm II}~\cite{s13,s16}, which presumably depends on the conjecture that
the supersymmetrization of spacetime might be unentangled with the introduction of the auxiliary worlvolume metric,
i.e. with the 2-metric prescription. However, it was argued~\cite{s10} that this would create considerable
algebraic complications\footnote{No such difficulties occur for super $p$-branes, see, for instance, ref.~\cite{s15}.}
if one attempted to introduce the auxiliary worldvolume metric field in the $Dp$-brane action of the Born-Infeld type.
We may say, it remains somehow ambiguous and needs further studies to construct the supersymmetric 3-metric actions of $Dp$-branes.

As to possible quantum theories of the actions with 3-metrics, we do not know how an exact role the double Weyl invariance plays
in the procedure of quantization, but do know that
it should not be useless because the Weyl invariance of the string action with an auxiliary worldsheet
metric has played a crucial role in the covariant quantization.
However, the $Dp$-brane case is much more
complicated than that of $p$-branes (strings when $p=1$). In fact, we are trying to make canonical Hamiltonian analyses for the simplest formulation
that possesses the double
Weyl invariance, i.e. the $D3$-brane action eq.~(33) as our first example on dealing with this problem, and will, if available, give the result
in a separate work.
Other questions related to the 3-metric actions, such as duality with respect to field strengths
and, on the other hand, construction of new conformal couplings in any worldvolume
dimension to the auxiliary scalar fields $\Phi(\xi)$ and $\Psi(\xi)$ promoted as dynamical variables~\cite{s30},
are also under consideration.

Finally, if we set the dilaton field and NS antisymmetric two-form field be zero and deal with metrics, gauge fields and field strengths
back to spacetime
instead of worldvolume in all series of $Dp$-brane actions derived above, we therefore obtain the results for the Born-Infeld theory.

\vspace{10mm}
\noindent
{\bf Acknowledgments}
\par
The author would like to thank H.J.W. M\"uller-Kirsten, N. Ohta, and L. Zhao for discussions.
He was indebted to the Associate Scheme provided by the Abdus Salam International Centre for
Theoretical Physics where part of the work was performed. This work was supported in part by the National Natural
Science Foundation of China under grants No.10675061 and No.10275052, by the Ministry of Education of China under grant No.20060055006,
and by Nankai University under grant No.J02013.

\newpage
\section*{Appendix A \hspace{.24cm}36 Forms of Series {\rm III} with Order (${\cal G}_{{\mu}{\nu}}$, $g_{{\mu}{\nu}}$)}
\begin{equation}
S^{{\rm III}}_{11}=-\frac{T_{p}}{16}\int d^{p+1}{\xi}e^{-\phi}(-{\gamma})^{1/4}(-{\omega})^{1/4}\Big[{\gamma}^{{\mu}{\nu}}{\cal G}_{{\mu}{\nu}}-(p-3)
\Big]\Big[{\omega}^{{\lambda}{\sigma}}g_{{\lambda}{\sigma}}-(p-3)\Big],
\end{equation}
\begin{equation}
S^{{\rm III}}_{12}=-\frac{T_{p}}{16}\int d^{p+1}{\xi}e^{-\phi}(-{\gamma})^{1/4}(-{\omega})^{1/4}\Big[{\gamma}^{{\mu}{\nu}}{\cal G}_{{\mu}{\nu}}-(p-3)
\Big]\Big[{\Psi}^{(p-3)/4}{\omega}^{{\lambda}{\sigma}}g_{{\lambda}{\sigma}}-(p-3){\Psi}^{(p+1)/4}\Big],
\end{equation}
\begin{equation}
S^{{\rm III}}_{13}=-\frac{T_{p}}{4}\int d^{p+1}{\xi}e^{-\phi}(-{\gamma})^{1/4}(-{\omega})^{1/4}\Big[{\gamma}^{{\mu}{\nu}}{\cal G}_{{\mu}{\nu}}-(p-3)
\Big]\left(\frac{1}{p+1}{\omega}^{{\lambda}{\sigma}}g_{{\lambda}{\sigma}}\right)^{(p+1)/4},
\end{equation}

\begin{equation}
S^{{\rm III}}_{1{\bar 1}}=-\frac{T_{p}}{16}\int d^{p+1}{\xi}e^{-\phi}(-{\gamma})^{1/4}(-{\omega})^{1/4}\Big[{\gamma}^{{\mu}{\nu}}{\cal G}_{{\mu}{\nu}}
-(p-3)\Big]\Big[-{\omega}_{{\lambda}{\sigma}}g^{{\lambda}{\sigma}}+(p+5)\Big],
\end{equation}
\begin{equation}
S^{{\rm III}}_{1{\bar 2}}=-\frac{T_{p}}{16}\int d^{p+1}{\xi}e^{-\phi}(-{\gamma})^{1/4}(-{\omega})^{1/4}\Big[{\gamma}^{{\mu}{\nu}}{\cal G}_{{\mu}{\nu}}
-(p-3)\Big]\Big[-{\Psi}^{(p+5)/4}{\omega}_{{\lambda}{\sigma}}g^{{\lambda}{\sigma}}+(p+5){\Psi}^{(p+1)/4}\Big],
\end{equation}
\begin{equation}
S^{{\rm III}}_{1{\bar 3}}=-\frac{T_{p}}{4}\int d^{p+1}{\xi}e^{-\phi}(-{\gamma})^{1/4}(-{\omega})^{1/4}\Big[{\gamma}^{{\mu}{\nu}}{\cal G}_{{\mu}{\nu}}
-(p-3)\Big]\left(\frac{1}{p+1}{\omega}_{{\lambda}{\sigma}}g^{{\lambda}{\sigma}}\right)^{-(p+1)/4},
\end{equation}

\begin{equation}
S^{{\rm III}}_{21}=-\frac{T_{p}}{16}\int d^{p+1}{\xi}e^{-\phi}(-{\gamma})^{1/4}(-{\omega})^{1/4}\Big[{\Phi}^{(p-3)/4}{\gamma}^{{\mu}{\nu}}
{\cal G}_{{\mu}{\nu}}-(p-3){\Phi}^{(p+1)/4}
\Big]\Big[{\omega}^{{\lambda}{\sigma}}g_{{\lambda}{\sigma}}-(p-3)\Big],
\end{equation}
\begin{eqnarray}
S^{{\rm III}}_{22}=-\frac{T_{p}}{16}\int &d^{p+1}{\xi}& e^{-\phi}(-{\gamma})^{1/4}(-{\omega})^{1/4}\Big[{\Phi}^{(p-3)/4}{\gamma}^{{\mu}{\nu}}
{\cal G}_{{\mu}{\nu}}-(p-3){\Phi}^{(p+1)/4}\Big]\nonumber\\
& & \times \Big[{\Psi}^{(p-3)/4}{\omega}^{{\lambda}{\sigma}}g_{{\lambda}{\sigma}}-(p-3){\Psi}^{(p+1)/4}\Big],
\end{eqnarray}
\begin{eqnarray}
S^{{\rm III}}_{23}=-\frac{T_{p}}{4}\int &d^{p+1}{\xi}& e^{-\phi}(-{\gamma})^{1/4}(-{\omega})^{1/4}\Big[{\Phi}^{(p-3)/4}{\gamma}^{{\mu}{\nu}}
{\cal G}_{{\mu}{\nu}}-(p-3){\Phi}^{(p+1)/4}\Big]\nonumber\\
& & \times \left(\frac{1}{p+1}{\omega}^{{\lambda}{\sigma}}g_{{\lambda}{\sigma}}\right)^{(p+1)/4},
\end{eqnarray}

\begin{equation}
S^{{\rm III}}_{2{\bar 1}}=-\frac{T_{p}}{16}\int d^{p+1}{\xi}e^{-\phi}(-{\gamma})^{1/4}(-{\omega})^{1/4}\Big[{\Phi}^{(p-3)/4}{\gamma}^{{\mu}{\nu}}
{\cal G}_{{\mu}{\nu}}-(p-3){\Phi}^{(p+1)/4}\Big]\Big[-{\omega}_{{\lambda}{\sigma}}g^{{\lambda}{\sigma}}+(p+5)\Big],
\end{equation}
\begin{eqnarray}
S^{{\rm III}}_{2{\bar 2}}=-\frac{T_{p}}{16}\int &d^{p+1}{\xi}& e^{-\phi}(-{\gamma})^{1/4}(-{\omega})^{1/4}\Big[{\Phi}^{(p-3)/4}{\gamma}^{{\mu}{\nu}}
{\cal G}_{{\mu}{\nu}}-(p-3){\Phi}^{(p+1)/4}\Big]\nonumber\\
& & \times \Big[-{\Psi}^{(p+5)/4}{\omega}_{{\lambda}{\sigma}}g^{{\lambda}{\sigma}}+(p+5){\Psi}^{(p+1)/4}\Big],
\end{eqnarray}
\begin{eqnarray}
S^{{\rm III}}_{2{\bar 3}}=-\frac{T_{p}}{4}\int &d^{p+1}{\xi}& e^{-\phi}(-{\gamma})^{1/4}(-{\omega})^{1/4}\Big[{\Phi}^{(p-3)/4}{\gamma}^{{\mu}{\nu}}
{\cal G}_{{\mu}{\nu}}-(p-3){\Phi}^{(p+1)/4}\Big]\nonumber\\
& & \times \left(\frac{1}{p+1}{\omega}_{{\lambda}{\sigma}}g^{{\lambda}{\sigma}}\right)^{-(p+1)/4},
\end{eqnarray}

\begin{equation}
S^{{\rm III}}_{31}=-\frac{T_{p}}{4}\int d^{p+1}{\xi}e^{-\phi}(-{\gamma})^{1/4}(-{\omega})^{1/4}\left(\frac{1}{p+1}{\gamma}^{{\mu}{\nu}}
{\cal G}_{{\mu}{\nu}}\right)^{(p+1)/4}\Big[{\omega}^{{\lambda}{\sigma}}g_{{\lambda}{\sigma}}-(p-3)\Big],
\end{equation}
\begin{eqnarray}
S^{{\rm III}}_{32}=-\frac{T_{p}}{4}\int &d^{p+1}{\xi}& e^{-\phi}(-{\gamma})^{1/4}(-{\omega})^{1/4}\left(\frac{1}{p+1}{\gamma}^{{\mu}{\nu}}
{\cal G}_{{\mu}{\nu}}\right)^{(p+1)/4}\nonumber\\
& & \times \Big[{\Psi}^{(p-3)/4}{\omega}^{{\lambda}{\sigma}}g_{{\lambda}{\sigma}}-(p-3){\Psi}^{(p+1)/4}\Big],
\end{eqnarray}
\begin{equation}
S^{{\rm III}}_{33}=-T_{p}\int d^{p+1}{\xi}e^{-\phi}(-{\gamma})^{1/4}(-{\omega})^{1/4}\left(\frac{1}{p+1}{\gamma}^{{\mu}{\nu}}
{\cal G}_{{\mu}{\nu}}\right)^{(p+1)/4}
\left(\frac{1}{p+1}{\omega}^{{\lambda}{\sigma}}g_{{\lambda}{\sigma}}\right)^{(p+1)/4},
\end{equation}

\begin{equation}
S^{{\rm III}}_{3{\bar 1}}=-\frac{T_{p}}{4}\int d^{p+1}{\xi}e^{-\phi}(-{\gamma})^{1/4}(-{\omega})^{1/4}\left(\frac{1}{p+1}{\gamma}^{{\mu}{\nu}}
{\cal G}_{{\mu}{\nu}}\right)^{(p+1)/4}
\Big[-{\omega}_{{\lambda}{\sigma}}g^{{\lambda}{\sigma}}+(p+5)\Big],
\end{equation}
\begin{eqnarray}
S^{{\rm III}}_{3{\bar 2}}=-\frac{T_{p}}{4}\int &d^{p+1}{\xi}& e^{-\phi}(-{\gamma})^{1/4}(-{\omega})^{1/4}\left(\frac{1}{p+1}{\gamma}^{{\mu}{\nu}}
{\cal G}_{{\mu}{\nu}}\right)^{(p+1)/4}\nonumber\\
& & \times \Big[-{\Psi}^{(p+5)/4}{\omega}_{{\lambda}{\sigma}}g^{{\lambda}{\sigma}}+(p+5){\Psi}^{(p+1)/4}\Big],
\end{eqnarray}
\begin{equation}
S^{{\rm III}}_{3{\bar 3}}=-T_{p}\int d^{p+1}{\xi}e^{-\phi}(-{\gamma})^{1/4}(-{\omega})^{1/4}\left(\frac{1}{p+1}{\gamma}^{{\mu}{\nu}}
{\cal G}_{{\mu}{\nu}}\right)^{(p+1)/4}
\left(\frac{1}{p+1}{\omega}_{{\lambda}{\sigma}}g^{{\lambda}{\sigma}}\right)^{-(p+1)/4},
\end{equation}

\begin{equation}
S^{{\rm III}}_{{\bar 1}1}=-\frac{T_{p}}{16}\int d^{p+1}{\xi}e^{-\phi}(-{\gamma})^{1/4}(-{\omega})^{1/4}\Big[{-\gamma}_{{\mu}{\nu}}{\cal G}^{{\mu}{\nu}}
+(p+5)\Big]\Big[{\omega}^{{\lambda}{\sigma}}g_{{\lambda}{\sigma}}-(p-3)\Big],
\end{equation}
\begin{equation}
S^{{\rm III}}_{{\bar 1}2}=-\frac{T_{p}}{16}\int d^{p+1}{\xi}e^{-\phi}(-{\gamma})^{1/4}(-{\omega})^{1/4}\Big[{-\gamma}_{{\mu}{\nu}}{\cal G}^{{\mu}{\nu}}
+(p+5)\Big]\Big[{\Psi}^{(p-3)/4}{\omega}^{{\lambda}{\sigma}}g_{{\lambda}{\sigma}}-(p-3){\Psi}^{(p+1)/4}\Big],
\end{equation}
\begin{equation}
S^{{\rm III}}_{{\bar 1}3}=-\frac{T_{p}}{4}\int d^{p+1}{\xi}e^{-\phi}(-{\gamma})^{1/4}(-{\omega})^{1/4}\Big[{-\gamma}_{{\mu}{\nu}}{\cal G}^{{\mu}{\nu}}
+(p+5)\Big]\left(\frac{1}{p+1}{\omega}^{{\lambda}{\sigma}}g_{{\lambda}{\sigma}}\right)^{(p+1)/4},
\end{equation}

\begin{equation}
S^{{\rm III}}_{{\bar 1}{\bar 1}}=-\frac{T_{p}}{16}\int d^{p+1}{\xi}e^{-\phi}(-{\gamma})^{1/4}(-{\omega})^{1/4}\Big[{-\gamma}_{{\mu}{\nu}}
{\cal G}^{{\mu}{\nu}}+(p+5)\Big]\Big[-{\omega}_{{\lambda}{\sigma}}g^{{\lambda}{\sigma}}+(p+5)\Big],
\end{equation}
\begin{equation}
S^{{\rm III}}_{{\bar 1}{\bar 2}}=-\frac{T_{p}}{16}\int d^{p+1}{\xi}e^{-\phi}(-{\gamma})^{1/4}(-{\omega})^{1/4}\Big[{-\gamma}_{{\mu}{\nu}}
{\cal G}^{{\mu}{\nu}}+(p+5)\Big]\Big[-{\Psi}^{(p+5)/4}{\omega}_{{\lambda}{\sigma}}g^{{\lambda}{\sigma}}+(p+5){\Psi}^{(p+1)/4}\Big],
\end{equation}
\begin{equation}
S^{{\rm III}}_{{\bar 1}{\bar 3}}=-\frac{T_{p}}{4}\int d^{p+1}{\xi}e^{-\phi}(-{\gamma})^{1/4}(-{\omega})^{1/4}\Big[{-\gamma}_{{\mu}{\nu}}
{\cal G}^{{\mu}{\nu}}+(p+5)\Big]\left(\frac{1}{p+1}{\omega}_{{\lambda}{\sigma}}g^{{\lambda}{\sigma}}\right)^{-(p+1)/4},
\end{equation}

\begin{equation}
S^{{\rm III}}_{{\bar 2}1}=-\frac{T_{p}}{16}\int d^{p+1}{\xi}e^{-\phi}(-{\gamma})^{1/4}(-{\omega})^{1/4}\Big[-{\Phi}^{(p+5)/4}{\gamma}_{{\mu}{\nu}}
{\cal G}^{{\mu}{\nu}}+(p+5){\Phi}^{(p+1)/4}
\Big]\Big[{\omega}^{{\lambda}{\sigma}}g_{{\lambda}{\sigma}}-(p-3)\Big],
\end{equation}
\begin{eqnarray}
S^{{\rm III}}_{{\bar 2}2}=-\frac{T_{p}}{16}\int &d^{p+1}{\xi}& e^{-\phi}(-{\gamma})^{1/4}(-{\omega})^{1/4}\Big[-{\Phi}^{(p+5)/4}{\gamma}_{{\mu}{\nu}}
{\cal G}^{{\mu}{\nu}}+(p+5){\Phi}^{(p+1)/4}\Big]\nonumber\\
& & \times \Big[{\Psi}^{(p-3)/4}{\omega}^{{\lambda}{\sigma}}g_{{\lambda}{\sigma}}-(p-3){\Psi}^{(p+1)/4}\Big],
\end{eqnarray}
\begin{eqnarray}
S^{{\rm III}}_{{\bar 2}3}=-\frac{T_{p}}{4}\int &d^{p+1}{\xi}& e^{-\phi}(-{\gamma})^{1/4}(-{\omega})^{1/4}\Big[-{\Phi}^{(p+5)/4}{\gamma}_{{\mu}{\nu}}
{\cal G}^{{\mu}{\nu}}+(p+5){\Phi}^{(p+1)/4}\Big]\nonumber\\
& & \times \left(\frac{1}{p+1}{\omega}^{{\lambda}{\sigma}}g_{{\lambda}{\sigma}}\right)^{(p+1)/4},
\end{eqnarray}

\begin{equation}
S^{{\rm III}}_{{\bar 2}{\bar 1}}=-\frac{T_{p}}{16}\int d^{p+1}{\xi}e^{-\phi}(-{\gamma})^{1/4}(-{\omega})^{1/4}\Big[-{\Phi}^{(p+5)/4}{\gamma}_{{\mu}{\nu}}
{\cal G}^{{\mu}{\nu}}+(p+5){\Phi}^{(p+1)/4}\Big]\Big[-{\omega}_{{\lambda}{\sigma}}g^{{\lambda}{\sigma}}+(p+5)\Big],
\end{equation}
\begin{eqnarray}
S^{{\rm III}}_{{\bar 2}{\bar 2}}=-\frac{T_{p}}{16}\int &d^{p+1}{\xi}& e^{-\phi}(-{\gamma})^{1/4}(-{\omega})^{1/4}\Big[-{\Phi}^{(p+5)/4}{\gamma}_{{\mu}
{\nu}}{\cal G}^{{\mu}{\nu}}+(p+5){\Phi}^{(p+1)/4}\Big]\nonumber\\
& & \times \Big[-{\Psi}^{(p+5)/4}{\omega}_{{\lambda}{\sigma}}g^{{\lambda}{\sigma}}+(p+5){\Psi}^{(p+1)/4}\Big],
\end{eqnarray}
\begin{eqnarray}
S^{{\rm III}}_{{\bar 2}{\bar 3}}=-\frac{T_{p}}{4}\int &d^{p+1}{\xi}& e^{-\phi}(-{\gamma})^{1/4}(-{\omega})^{1/4}\Big[-{\Phi}^{(p+5)/4}{\gamma}_{{\mu}
{\nu}}{\cal G}^{{\mu}{\nu}}+(p+5){\Phi}^{(p+1)/4}\Big]\nonumber\\
& & \times \left(\frac{1}{p+1}{\omega}_{{\lambda}{\sigma}}g^{{\lambda}{\sigma}}\right)^{-(p+1)/4},
\end{eqnarray}

\begin{equation}
S^{{\rm III}}_{{\bar 3}1}=-\frac{T_{p}}{4}\int d^{p+1}{\xi}e^{-\phi}(-{\gamma})^{1/4}(-{\omega})^{1/4}\left(\frac{1}{p+1}{\gamma}_{{\mu}{\nu}}
{\cal G}^{{\mu}{\nu}}
\right)^{-(p+1)/4}\Big[{\omega}^{{\lambda}{\sigma}}g_{{\lambda}{\sigma}}-(p-3)\Big],
\end{equation}
\begin{eqnarray}
S^{{\rm III}}_{{\bar 3}2}=-\frac{T_{p}}{4}\int &d^{p+1}{\xi}& e^{-\phi}(-{\gamma})^{1/4}(-{\omega})^{1/4}\left(\frac{1}{p+1}{\gamma}_{{\mu}{\nu}}
{\cal G}^{{\mu}{\nu}}\right)^{-(p+1)/4}\nonumber\\
& & \times \Big[{\Psi}^{(p-3)/4}{\omega}^{{\lambda}{\sigma}}g_{{\lambda}{\sigma}}-(p-3){\Psi}^{(p+1)/4}\Big],
\end{eqnarray}
\begin{equation}
S^{{\rm III}}_{{\bar 3}3}=-T_{p}\int d^{p+1}{\xi}e^{-\phi}(-{\gamma})^{1/4}(-{\omega})^{1/4}\left(\frac{1}{p+1}{\gamma}_{{\mu}{\nu}}{\cal G}^{{\mu}{\nu}}
\right)^{-(p+1)/4}
\left(\frac{1}{p+1}{\omega}^{{\lambda}{\sigma}}g_{{\lambda}{\sigma}}\right)^{(p+1)/4},
\end{equation}

\begin{equation}
S^{{\rm III}}_{{\bar 3}{\bar 1}}=-\frac{T_{p}}{4}\int d^{p+1}{\xi}e^{-\phi}(-{\gamma})^{1/4}(-{\omega})^{1/4}\left(\frac{1}{p+1}{\gamma}_{{\mu}{\nu}}
{\cal G}^{{\mu}{\nu}}\right)^{-(p+1)/4}
\Big[-{\omega}_{{\lambda}{\sigma}}g^{{\lambda}{\sigma}}+(p+5)\Big],
\end{equation}
\begin{eqnarray}
S^{{\rm III}}_{{\bar 3}{\bar 2}}=-\frac{T_{p}}{4}\int &d^{p+1}{\xi}& e^{-\phi}(-{\gamma})^{1/4}(-{\omega})^{1/4}\left(\frac{1}{p+1}{\gamma}_{{\mu}{\nu}}
{\cal G}^{{\mu}{\nu}}\right)^{-(p+1)/4}\nonumber\\
& & \times \Big[-{\Psi}^{(p+5)/4}{\omega}_{{\lambda}{\sigma}}g^{{\lambda}{\sigma}}+(p+5){\Psi}^{(p+1)/4}\Big],
\end{eqnarray}
\begin{equation}
S^{{\rm III}}_{{\bar 3}{\bar 3}}=-T_{p}\int d^{p+1}{\xi}e^{-\phi}(-{\gamma})^{1/4}(-{\omega})^{1/4}\left(\frac{1}{p+1}{\gamma}_{{\mu}{\nu}}
{\cal G}^{{\mu}{\nu}}\right)^{-(p+1)/4}
\left(\frac{1}{p+1}{\omega}_{{\lambda}{\sigma}}g^{{\lambda}{\sigma}}\right)^{-(p+1)/4}.
\end{equation}

\section*{Appendix B \hspace{.24cm}36 Forms of Series {\rm III} with Order ($g_{{\mu}{\nu}}$, ${\cal G}_{{\mu}{\nu}}$)}
\begin{equation}
{S^{\prime}}^{{\rm III}}_{11}=-\frac{T_{p}}{16}\int d^{p+1}{\xi}e^{-\phi}(-{\gamma})^{1/4}(-{\omega})^{1/4}\Big[{\gamma}^{{\mu}{\nu}}
g_{{\mu}{\nu}}-(p-3)
\Big]\Big[{\omega}^{{\lambda}{\sigma}}{\cal G}_{{\lambda}{\sigma}}-(p-3)\Big],
\end{equation}
\begin{equation}
{S^{\prime}}^{{\rm III}}_{12}=-\frac{T_{p}}{16}\int d^{p+1}{\xi}e^{-\phi}(-{\gamma})^{1/4}(-{\omega})^{1/4}\Big[{\gamma}^{{\mu}{\nu}}
g_{{\mu}{\nu}}-(p-3)
\Big]\Big[{\Psi}^{(p-3)/4}{\omega}^{{\lambda}{\sigma}}{\cal G}_{{\lambda}{\sigma}}-(p-3){\Psi}^{(p+1)/4}\Big],
\end{equation}
\begin{equation}
{S^{\prime}}^{{\rm III}}_{13}=-\frac{T_{p}}{4}\int d^{p+1}{\xi}e^{-\phi}(-{\gamma})^{1/4}(-{\omega})^{1/4}\Big[{\gamma}^{{\mu}{\nu}}
g_{{\mu}{\nu}}-(p-3)
\Big]\left(\frac{1}{p+1}{\omega}^{{\lambda}{\sigma}}{\cal G}_{{\lambda}{\sigma}}\right)^{(p+1)/4},
\end{equation}

\begin{equation}
{S^{\prime}}^{{\rm III}}_{1{\bar 1}}=-\frac{T_{p}}{16}\int d^{p+1}{\xi}e^{-\phi}(-{\gamma})^{1/4}(-{\omega})^{1/4}\Big[{\gamma}^{{\mu}{\nu}}
g_{{\mu}{\nu}}
-(p-3)\Big]\Big[-{\omega}_{{\lambda}{\sigma}}{\cal G}^{{\lambda}{\sigma}}+(p+5)\Big],
\end{equation}
\begin{equation}
{S^{\prime}}^{{\rm III}}_{1{\bar 2}}=-\frac{T_{p}}{16}\int d^{p+1}{\xi}e^{-\phi}(-{\gamma})^{1/4}(-{\omega})^{1/4}\Big[{\gamma}^{{\mu}{\nu}}
g_{{\mu}{\nu}}
-(p-3)\Big]\Big[-{\Psi}^{(p+5)/4}{\omega}_{{\lambda}{\sigma}}{\cal G}^{{\lambda}{\sigma}}+(p+5){\Psi}^{(p+1)/4}\Big],
\end{equation}
\begin{equation}
{S^{\prime}}^{{\rm III}}_{1{\bar 3}}=-\frac{T_{p}}{4}\int d^{p+1}{\xi}e^{-\phi}(-{\gamma})^{1/4}(-{\omega})^{1/4}\Big[{\gamma}^{{\mu}{\nu}}
g_{{\mu}{\nu}}
-(p-3)\Big]\left(\frac{1}{p+1}{\omega}_{{\lambda}{\sigma}}{\cal G}^{{\lambda}{\sigma}}\right)^{-(p+1)/4},
\end{equation}

\begin{equation}
{S^{\prime}}^{{\rm III}}_{21}=-\frac{T_{p}}{16}\int d^{p+1}{\xi}e^{-\phi}(-{\gamma})^{1/4}(-{\omega})^{1/4}\Big[{\Phi}^{(p-3)/4}{\gamma}^{{\mu}{\nu}}
g_{{\mu}{\nu}}-(p-3){\Phi}^{(p+1)/4}
\Big]\Big[{\omega}^{{\lambda}{\sigma}}{\cal G}_{{\lambda}{\sigma}}-(p-3)\Big],
\end{equation}
\begin{eqnarray}
{S^{\prime}}^{{\rm III}}_{22}=-\frac{T_{p}}{16}\int &d^{p+1}{\xi}& e^{-\phi}(-{\gamma})^{1/4}(-{\omega})^{1/4}\Big[{\Phi}^{(p-3)/4}{\gamma}^{{\mu}{\nu}}
g_{{\mu}{\nu}}-(p-3){\Phi}^{(p+1)/4}\Big]\nonumber\\
& & \times \Big[{\Psi}^{(p-3)/4}{\omega}^{{\lambda}{\sigma}}{\cal G}_{{\lambda}{\sigma}}-(p-3){\Psi}^{(p+1)/4}\Big],
\end{eqnarray}
\begin{eqnarray}
{S^{\prime}}^{{\rm III}}_{23}=-\frac{T_{p}}{4}\int &d^{p+1}{\xi}& e^{-\phi}(-{\gamma})^{1/4}(-{\omega})^{1/4}\Big[{\Phi}^{(p-3)/4}{\gamma}^{{\mu}{\nu}}
g_{{\mu}{\nu}}-(p-3){\Phi}^{(p+1)/4}\Big]\nonumber\\
& & \times \left(\frac{1}{p+1}{\omega}^{{\lambda}{\sigma}}{\cal G}_{{\lambda}{\sigma}}\right)^{(p+1)/4},
\end{eqnarray}

\begin{equation}
{S^{\prime}}^{{\rm III}}_{2{\bar 1}}=-\frac{T_{p}}{16}\int d^{p+1}{\xi}e^{-\phi}(-{\gamma})^{1/4}(-{\omega})^{1/4}\Big[{\Phi}^{(p-3)/4}
{\gamma}^{{\mu}{\nu}}
g_{{\mu}{\nu}}-(p-3){\Phi}^{(p+1)/4}\Big]\Big[-{\omega}_{{\lambda}{\sigma}}{\cal G}^{{\lambda}{\sigma}}+(p+5)\Big],
\end{equation}
\begin{eqnarray}
{S^{\prime}}^{{\rm III}}_{2{\bar 2}}=-\frac{T_{p}}{16}\int &d^{p+1}{\xi}& e^{-\phi}(-{\gamma})^{1/4}(-{\omega})^{1/4}\Big[{\Phi}^{(p-3)/4}
{\gamma}^{{\mu}{\nu}}
g_{{\mu}{\nu}}-(p-3){\Phi}^{(p+1)/4}\Big]\nonumber\\
& & \times \Big[-{\Psi}^{(p+5)/4}{\omega}_{{\lambda}{\sigma}}{\cal G}^{{\lambda}{\sigma}}+(p+5){\Psi}^{(p+1)/4}\Big],
\end{eqnarray}
\begin{eqnarray}
{S^{\prime}}^{{\rm III}}_{2{\bar 3}}=-\frac{T_{p}}{4}\int &d^{p+1}{\xi}& e^{-\phi}(-{\gamma})^{1/4}(-{\omega})^{1/4}\Big[{\Phi}^{(p-3)/4}
{\gamma}^{{\mu}{\nu}}
g_{{\mu}{\nu}}-(p-3){\Phi}^{(p+1)/4}\Big]\nonumber\\
& & \times \left(\frac{1}{p+1}{\omega}_{{\lambda}{\sigma}}{\cal G}^{{\lambda}{\sigma}}\right)^{-(p+1)/4},
\end{eqnarray}

\begin{equation}
{S^{\prime}}^{{\rm III}}_{31}=-\frac{T_{p}}{4}\int d^{p+1}{\xi}e^{-\phi}(-{\gamma})^{1/4}(-{\omega})^{1/4}\left(\frac{1}{p+1}{\gamma}^{{\mu}{\nu}}
g_{{\mu}{\nu}}\right)^{(p+1)/4}\Big[{\omega}^{{\lambda}{\sigma}}{\cal G}_{{\lambda}{\sigma}}-(p-3)\Big],
\end{equation}
\begin{eqnarray}
{S^{\prime}}^{{\rm III}}_{32}=-\frac{T_{p}}{4}\int &d^{p+1}{\xi}& e^{-\phi}(-{\gamma})^{1/4}(-{\omega})^{1/4}\left(\frac{1}{p+1}{\gamma}^{{\mu}{\nu}}
g_{{\mu}{\nu}}\right)^{(p+1)/4}\nonumber\\
& & \times \Big[{\Psi}^{(p-3)/4}{\omega}^{{\lambda}{\sigma}}{\cal G}_{{\lambda}{\sigma}}-(p-3){\Psi}^{(p+1)/4}\Big],
\end{eqnarray}
\begin{equation}
{S^{\prime}}^{{\rm III}}_{33}=-T_{p}\int d^{p+1}{\xi}e^{-\phi}(-{\gamma})^{1/4}(-{\omega})^{1/4}\left(\frac{1}{p+1}{\gamma}^{{\mu}{\nu}}
g_{{\mu}{\nu}}\right)^{(p+1)/4}
\left(\frac{1}{p+1}{\omega}^{{\lambda}{\sigma}}{\cal G}_{{\lambda}{\sigma}}\right)^{(p+1)/4},
\end{equation}

\begin{equation}
{S^{\prime}}^{{\rm III}}_{3{\bar 1}}=-\frac{T_{p}}{4}\int d^{p+1}{\xi}e^{-\phi}(-{\gamma})^{1/4}(-{\omega})^{1/4}\left(\frac{1}{p+1}{\gamma}^{{\mu}{\nu}}
g_{{\mu}{\nu}}\right)^{(p+1)/4}
\Big[-{\omega}_{{\lambda}{\sigma}}{\cal G}^{{\lambda}{\sigma}}+(p+5)\Big],
\end{equation}
\begin{eqnarray}
{S^{\prime}}^{{\rm III}}_{3{\bar 2}}=-\frac{T_{p}}{4}\int &d^{p+1}{\xi}& e^{-\phi}(-{\gamma})^{1/4}(-{\omega})^{1/4}\left(\frac{1}{p+1}
{\gamma}^{{\mu}{\nu}}
g_{{\mu}{\nu}}\right)^{(p+1)/4}\nonumber\\
& & \times \Big[-{\Psi}^{(p+5)/4}{\omega}_{{\lambda}{\sigma}}{\cal G}^{{\lambda}{\sigma}}+(p+5){\Psi}^{(p+1)/4}\Big],
\end{eqnarray}
\begin{equation}
{S^{\prime}}^{{\rm III}}_{3{\bar 3}}=-T_{p}\int d^{p+1}{\xi}e^{-\phi}(-{\gamma})^{1/4}(-{\omega})^{1/4}\left(\frac{1}{p+1}{\gamma}^{{\mu}{\nu}}
g_{{\mu}{\nu}}\right)^{(p+1)/4}
\left(\frac{1}{p+1}{\omega}_{{\lambda}{\sigma}}{\cal G}^{{\lambda}{\sigma}}\right)^{-(p+1)/4},
\end{equation}

\begin{equation}
{S^{\prime}}^{{\rm III}}_{{\bar 1}1}=-\frac{T_{p}}{16}\int d^{p+1}{\xi}e^{-\phi}(-{\gamma})^{1/4}(-{\omega})^{1/4}\Big[{-\gamma}_{{\mu}{\nu}}
g^{{\mu}{\nu}}
+(p+5)\Big]\Big[{\omega}^{{\lambda}{\sigma}}{\cal G}_{{\lambda}{\sigma}}-(p-3)\Big],
\end{equation}
\begin{equation}
{S^{\prime}}^{{\rm III}}_{{\bar 1}2}=-\frac{T_{p}}{16}\int d^{p+1}{\xi}e^{-\phi}(-{\gamma})^{1/4}(-{\omega})^{1/4}\Big[{-\gamma}_{{\mu}{\nu}}
g^{{\mu}{\nu}}
+(p+5)\Big]\Big[{\Psi}^{(p-3)/4}{\omega}^{{\lambda}{\sigma}}{\cal G}_{{\lambda}{\sigma}}-(p-3){\Psi}^{(p+1)/4}\Big],
\end{equation}
\begin{equation}
{S^{\prime}}^{{\rm III}}_{{\bar 1}3}=-\frac{T_{p}}{4}\int d^{p+1}{\xi}e^{-\phi}(-{\gamma})^{1/4}(-{\omega})^{1/4}\Big[{-\gamma}_{{\mu}{\nu}}
g^{{\mu}{\nu}}
+(p+5)\Big]\left(\frac{1}{p+1}{\omega}^{{\lambda}{\sigma}}{\cal G}_{{\lambda}{\sigma}}\right)^{(p+1)/4},
\end{equation}

\begin{equation}
{S^{\prime}}^{{\rm III}}_{{\bar 1}{\bar 1}}=-\frac{T_{p}}{16}\int d^{p+1}{\xi}e^{-\phi}(-{\gamma})^{1/4}(-{\omega})^{1/4}\Big[{-\gamma}_{{\mu}{\nu}}
g^{{\mu}{\nu}}+(p+5)\Big]\Big[-{\omega}_{{\lambda}{\sigma}}{\cal G}^{{\lambda}{\sigma}}+(p+5)\Big],
\end{equation}
\begin{equation}
{S^{\prime}}^{{\rm III}}_{{\bar 1}{\bar 2}}=-\frac{T_{p}}{16}\int d^{p+1}{\xi}e^{-\phi}(-{\gamma})^{1/4}(-{\omega})^{1/4}\Big[{-\gamma}_{{\mu}{\nu}}
g^{{\mu}{\nu}}+(p+5)\Big]\Big[-{\Psi}^{(p+5)/4}{\omega}_{{\lambda}{\sigma}}{\cal G}^{{\lambda}{\sigma}}+(p+5){\Psi}^{(p+1)/4}\Big],
\end{equation}
\begin{equation}
{S^{\prime}}^{{\rm III}}_{{\bar 1}{\bar 3}}=-\frac{T_{p}}{4}\int d^{p+1}{\xi}e^{-\phi}(-{\gamma})^{1/4}(-{\omega})^{1/4}\Big[{-\gamma}_{{\mu}{\nu}}
g^{{\mu}{\nu}}+(p+5)\Big]\left(\frac{1}{p+1}{\omega}_{{\lambda}{\sigma}}{\cal G}^{{\lambda}{\sigma}}\right)^{-(p+1)/4},
\end{equation}

\begin{equation}
{S^{\prime}}^{{\rm III}}_{{\bar 2}1}=-\frac{T_{p}}{16}\int d^{p+1}{\xi}e^{-\phi}(-{\gamma})^{1/4}(-{\omega})^{1/4}\Big[-{\Phi}^{(p+5)/4}
{\gamma}_{{\mu}{\nu}}
g^{{\mu}{\nu}}+(p+5){\Phi}^{(p+1)/4}
\Big]\Big[{\omega}^{{\lambda}{\sigma}}{\cal G}_{{\lambda}{\sigma}}-(p-3)\Big],
\end{equation}
\begin{eqnarray}
{S^{\prime}}^{{\rm III}}_{{\bar 2}2}=-\frac{T_{p}}{16}\int &d^{p+1}{\xi}& e^{-\phi}(-{\gamma})^{1/4}(-{\omega})^{1/4}\Big[-{\Phi}^{(p+5)/4}
{\gamma}_{{\mu}{\nu}}
g^{{\mu}{\nu}}+(p+5){\Phi}^{(p+1)/4}\Big]\nonumber\\
& & \times \Big[{\Psi}^{(p-3)/4}{\omega}^{{\lambda}{\sigma}}{\cal G}_{{\lambda}{\sigma}}-(p-3){\Psi}^{(p+1)/4}\Big],
\end{eqnarray}
\begin{eqnarray}
{S^{\prime}}^{{\rm III}}_{{\bar 2}3}=-\frac{T_{p}}{4}\int &d^{p+1}{\xi}& e^{-\phi}(-{\gamma})^{1/4}(-{\omega})^{1/4}\Big[-{\Phi}^{(p+5)/4}
{\gamma}_{{\mu}{\nu}}
g^{{\mu}{\nu}}+(p+5){\Phi}^{(p+1)/4}\Big]\nonumber\\
& & \times \left(\frac{1}{p+1}{\omega}^{{\lambda}{\sigma}}{\cal G}_{{\lambda}{\sigma}}\right)^{(p+1)/4},
\end{eqnarray}

\begin{equation}
{S^{\prime}}^{{\rm III}}_{{\bar 2}{\bar 1}}=-\frac{T_{p}}{16}\int d^{p+1}{\xi}e^{-\phi}(-{\gamma})^{1/4}(-{\omega})^{1/4}\Big[-{\Phi}^{(p+5)/4}
{\gamma}_{{\mu}{\nu}}
g^{{\mu}{\nu}}+(p+5){\Phi}^{(p+1)/4}\Big]\Big[-{\omega}_{{\lambda}{\sigma}}{\cal G}^{{\lambda}{\sigma}}+(p+5)\Big],
\end{equation}
\begin{eqnarray}
{S^{\prime}}^{{\rm III}}_{{\bar 2}{\bar 2}}=-\frac{T_{p}}{16}\int &d^{p+1}{\xi}& e^{-\phi}(-{\gamma})^{1/4}(-{\omega})^{1/4}\Big[-{\Phi}^{(p+5)/4}
{\gamma}_{{\mu}
{\nu}}g^{{\mu}{\nu}}+(p+5){\Phi}^{(p+1)/4}\Big]\nonumber\\
& & \times \Big[-{\Psi}^{(p+5)/4}{\omega}_{{\lambda}{\sigma}}{\cal G}^{{\lambda}{\sigma}}+(p+5){\Psi}^{(p+1)/4}\Big],
\end{eqnarray}
\begin{eqnarray}
{S^{\prime}}^{{\rm III}}_{{\bar 2}{\bar 3}}=-\frac{T_{p}}{4}\int &d^{p+1}{\xi}& e^{-\phi}(-{\gamma})^{1/4}(-{\omega})^{1/4}\Big[-{\Phi}^{(p+5)/4}
{\gamma}_{{\mu}
{\nu}}g^{{\mu}{\nu}}+(p+5){\Phi}^{(p+1)/4}\Big]\nonumber\\
& & \times \left(\frac{1}{p+1}{\omega}_{{\lambda}{\sigma}}{\cal G}^{{\lambda}{\sigma}}\right)^{-(p+1)/4},
\end{eqnarray}

\begin{equation}
{S^{\prime}}^{{\rm III}}_{{\bar 3}1}=-\frac{T_{p}}{4}\int d^{p+1}{\xi}e^{-\phi}(-{\gamma})^{1/4}(-{\omega})^{1/4}\left(\frac{1}{p+1}{\gamma}_{{\mu}{\nu}}
g^{{\mu}{\nu}}
\right)^{-(p+1)/4}\Big[{\omega}^{{\lambda}{\sigma}}{\cal G}_{{\lambda}{\sigma}}-(p-3)\Big],
\end{equation}
\begin{eqnarray}
{S^{\prime}}^{{\rm III}}_{{\bar 3}2}=-\frac{T_{p}}{4}\int &d^{p+1}{\xi}& e^{-\phi}(-{\gamma})^{1/4}(-{\omega})^{1/4}\left(\frac{1}{p+1}
{\gamma}_{{\mu}{\nu}}
g^{{\mu}{\nu}}\right)^{-(p+1)/4}\nonumber\\
& & \times \Big[{\Psi}^{(p-3)/4}{\omega}^{{\lambda}{\sigma}}{\cal G}_{{\lambda}{\sigma}}-(p-3){\Psi}^{(p+1)/4}\Big],
\end{eqnarray}
\begin{equation}
{S^{\prime}}^{{\rm III}}_{{\bar 3}3}=-T_{p}\int d^{p+1}{\xi}e^{-\phi}(-{\gamma})^{1/4}(-{\omega})^{1/4}\left(\frac{1}{p+1}{\gamma}_{{\mu}{\nu}}
g^{{\mu}{\nu}}
\right)^{-(p+1)/4}
\left(\frac{1}{p+1}{\omega}^{{\lambda}{\sigma}}{\cal G}_{{\lambda}{\sigma}}\right)^{(p+1)/4},
\end{equation}

\begin{equation}
{S^{\prime}}^{{\rm III}}_{{\bar 3}{\bar 1}}=-\frac{T_{p}}{4}\int d^{p+1}{\xi}e^{-\phi}(-{\gamma})^{1/4}(-{\omega})^{1/4}\left(\frac{1}{p+1}
{\gamma}_{{\mu}{\nu}}
g^{{\mu}{\nu}}\right)^{-(p+1)/4}
\Big[-{\omega}_{{\lambda}{\sigma}}{\cal G}^{{\lambda}{\sigma}}+(p+5)\Big],
\end{equation}
\begin{eqnarray}
{S^{\prime}}^{{\rm III}}_{{\bar 3}{\bar 2}}=-\frac{T_{p}}{4}\int &d^{p+1}{\xi}& e^{-\phi}(-{\gamma})^{1/4}(-{\omega})^{1/4}\left(\frac{1}{p+1}
{\gamma}_{{\mu}{\nu}}
g^{{\mu}{\nu}}\right)^{-(p+1)/4}\nonumber\\
& & \times \Big[-{\Psi}^{(p+5)/4}{\omega}_{{\lambda}{\sigma}}{\cal G}^{{\lambda}{\sigma}}+(p+5){\Psi}^{(p+1)/4}\Big],
\end{eqnarray}
\begin{equation}
{S^{\prime}}^{{\rm III}}_{{\bar 3}{\bar 3}}=-T_{p}\int d^{p+1}{\xi}e^{-\phi}(-{\gamma})^{1/4}(-{\omega})^{1/4}\left(\frac{1}{p+1}{\gamma}_{{\mu}{\nu}}
g^{{\mu}{\nu}}\right)^{-(p+1)/4}
\left(\frac{1}{p+1}{\omega}_{{\lambda}{\sigma}}{\cal G}^{{\lambda}{\sigma}}\right)^{-(p+1)/4}.
\end{equation}

\end{document}